\documentclass{aa}
\usepackage{lineno}
\usepackage[breaklinks=true]{hyperref}
\hypersetup{colorlinks,linkcolor=blue,citecolor=blue,urlcolor=blue}
\usepackage{natbib}
\bibpunct{(}{)}{;}{a}{}{,}
\usepackage{graphicx}
\usepackage{txfonts}
\usepackage{lscape}
\usepackage{longtable}
\usepackage{multirow}
\usepackage{multicol}
\usepackage{amssymb}
\usepackage{pifont}
\usepackage{geometry}
\usepackage{pdflscape}
\begin{document}

\title{Investigation of the stellar content in the IRAS\,05168+3634 star-forming region}

   \author{          N.M. Azatyan         }

   \institute{Byurakan Astrophysical Observatory, 0213, Aragatsotn prov., Armenia\\
              \email{nayazatyan@bao.sci.am}}

   \date{Received ; accepted }

  \abstract
 {}
   {We report the investigation results of the structure and content of a molecular cloud surrounding the source IRAS\,05168+3634 (also known as Mol\,9).}
   {We present a photometric analysis using the data of J, H, K UKIDSS, [3.6], [4.5]\,$\mu$m $Spitzer$-IRAC and 3.4, 4.6, 12, 22\,$\mu$m WISE databases. A multi-color criteria was used to identify the candidates of young stellar objects (YSOs) in the molecular cloud; in addition to IRAS\,05168+3634, there are four IRAS sources embedded in the same molecular cloud. Color-magnitude diagrams and the K luminosity function (KLF) were used to determine the basic parameters of stellar objects (spectral classes, masses, ages). To study the YSOs with longer wavelength photometry the radiative transfer models were used.}
   {Based on color-color and color-magnitude diagrams, we identified a rich population of embedded YSO candidates with infrared excess (Class\,0/I and Class\,II) and their characteristics in a quite large molecular cloud located in a region of 24\,arcmin radius. The molecular cloud includes 240 candidates of YSOs within the radii of subregions around five IRAS sources. The local distribution of identified YSOs in the molecular cloud frequently shows elongation and subclustering. The observed young subregions and parental molecular cloud morphologies are similar, especially when only the youngest Class\,I/0 sources are considered. The color-magnitude diagrams of the subregions suggest a very young stellar population. We construct the KLF of the subregions except for the IRAS\,05162+3639 region) and it shows unusually low values for $\alpha$ slope: 0.12--0.21. According to the values of the slopes of the KLFs, the age of the subregions can be estimated at 0.1--3\,Myr. The spectral energy distributions (SEDs) are constructed for 45 Class\,I and 75 Class\,II evolutionary stage YSOs and the received parameters of these YSOs are well correlated with the results obtained by other methods.}
  {}


   \maketitle
\section{Introduction}
\label{1}
Young embedded stellar clusters are important laboratories for understanding star formation process and early stellar evolution because they contain a statistically significant sample of stars formed from the same parental interstellar matter. The development of sensitive, large-format imaging arrays at near- and mid-infrared (NIR and MIR), submillimeter, and radio wavelengths has made it possible to obtain statistically significant and complete sampling of young embedded clusters within molecular clouds.

The widely accepted scenario of low- and intermediate-mass star formation is that they are formed by gravitational collapse, and the subsequent accretion of their parent molecular clouds and driving collimated outflows. However, the main mechanism leading to the formation of massive stars is debated as to whether it is disk accretion similar to that for lower mass stars \citep{yorke02}. Unlike low-mass stars, the high-mass star evolution is a fast process that lasts for no more than about 10$^{6}$ yr including the main sequence. During their formation, massive stars usually stay within their birth sites in clustered environments \citep{xin08}.

In this paper, we present a detailed study of a star-forming region in the vicinity of  IRAS\,05168+3634, which is also known as Mol\,9 in the catalogue of \citet{molinari96}. Within a 2\,arcmin radius of IRAS\,05168+3634, three objects have been detected with magnitudes of the  Midcourse Space Experiment (MSX) survey \citep{egan03}, one of which is associated with IRAS\,05168+3634. \citet{zhang05} have discovered a molecular outflow in this region.

IRAS\,05168+3634 is a luminous young stellar object (YSO) (longitude = 170.657, latitude = -00.27) with estimated L=24\,$\times$\,$10^3$\,$L_{\odot}$ \citep{varricatt10}; it is located in a high-mass star-forming region in the pre-UC\,HII phase \citep{wang09}. Compact H\,II regions and H$_{2}$O maser are sites of ongoing high-mass star formation which have not completely disrupted the surrounding dense, molecular gas. Thus, we have an opportunity to examine the stellar population in its initial configuration, before the rapid dynamical evolution which may occur once the molecular gas is cleared by stellar winds and the expanding region \citep{lada84}. There have been various detections in the region: H$_{2}$O maser emission \citep{zhang05}, NH$_3$ maser emission \citep{molinari96}, CS emission \citep{bronfman96}, a new detection of 44\,GHz CH$_{3}$OH methanol maser emission \citep{fontani10}, the  SiO (J = 2--1) line \citep{harju98}, the main lines at 1665\,MHz and 1667\,MHz OH maser \citep{ruiz16}, and four $^{13}$CO cores \citep{guan08}. No radio source has been associated with  IRAS\,05168+3634. \citet{molinari98} have detected 6\,cm radio emission 102\,arcsec away from IRAS\,05168+3634.

The embedded stellar cluster in this region was detected in the NIR and MIR by various authors \citep{kumar06,faustini09,azatyan16}. In \citet{azatyan16} it was shown that this is a bimodal cluster with 1.5\,arcmin radius from geometric center of the cluster that does not coincide with IRAS\,05168+3634. One of the subgroups is concentrated around IRAS\,05168+3634 and it should be noted that it does not contain a rich population compared to other concentrations.

The distance estimations of this region are different. A kinematic distance was estimated of 6.08\,kpc \citep{molinari96} based on systemic local standard of rest (LSR) velocity V$_{LSR}$ = $-$15.5$\pm$1.9\,km/s. The trigonometric parallax of IRAS\,05168+3634 with VERA is 0.532$\pm$0.053\,mas, which corresponds to a distance of $1.88^{+0.21}_{-0.17}$\,kpc, placing the region in the Perseus arm rather than the Outer arm \citep{sakai12}. This large difference of estimated distances causes some significant differences in physical parameters for individual members.

In the present paper, we tabulate of the YSO candidates in this star-forming region and discuss their spatial distribution. The observational data used to make the subsequent analysis are discussed in Section \ref{2}. The results and discussion are given in Section \ref{3}, and Section \ref{4} concludes with the results of this work.

\section{Observational data}
\label{2}
For our study we used the data covering a wide infrared (IR) range from NIR to far-infrared (FIR) wavelengths. The first is the archival NIR photometric data and images in the J, H, and K bands of the Galactic Plane Survey DR6 \citep[GPS,][]{lucas08} with a resolution of 0.1\,\arcsec\,/px, which is one of the five surveys of the UKIRT Infrared Deep Sky Survey (UKIDSS). UKIDSS GPS is complete to approximately 18.05 K\,mag, and for individual objects provides a probability (in percent) of being a star, galaxy, and noise accessible in VizieR service. 

Mid-IR wavelengths are ideal for studying disks because these wavelengths are well removed from the peak of the underlying stellar energy distribution, resulting in much greater excesses over the underlying stellar photosphere. Archival MIR observations of this region were obtained using the $Spitzer$ Space Telescope under the Galactic Legacy Infrared Midplane Survey Extraordinaire 360 (GLIMPSE\,360) program \citep{churchwell09}. GLIMPSE\,360 observations were taken using the InfraRed Array Camera \citep[IRAC,][]{fazio04} in 3.6, 4.5\,$\mu$m bands with a resolution of 0.6\,\arcsec\,/px. The point source photometric data in this star-forming region were downloaded from the NASA/IPAC Infrared Science Archive. We also used the Wide-field Infrared Survey Explorer \citep[WISE,][]{wright10} (3.4\,$\mu$m, 4.6\,$\mu$m, 12\,$\mu$m, and 22\,$\mu$m) and the MSX surveys (8.28\,$\mu$m, 12.13\,$\mu$m, 14.65\,$\mu$m, and 21.3\,$\mu$m) accessible in VizieR.

Since high-mass star formation happens deeply embedded, only observations at FIR and longer wavelengths can penetrate inside, and therefore provide essential information about such star-forming regions. This information may be provided by FIR wavelengths, in the range 70--500\,$\mu$m, obtained by using the Photodetector Array Camera and Spectrometer \citep[PACS,][]{poglitsch10} and  the Spectral and Photometric Imaging Receiver \citep[SPIRE,][]{griffin10} on the 3.5\,m $Herschel$ Space Observatory \citep{pilbratt10}. For our analyses, we used PACS 70, 160\,$\mu$m and  SPIRE 250, 350, and 500\,$\mu$m photometric data and images with resolutions varying from $\backsim$5.5\,\arcsec\, to 36\,\arcsec\,. The point and extended source photometry were downloaded from the NASA/IPAC Infrared Science Archive. $Herschel$ PACS data have better resolution with respect to the IRAS mission data; for this reason we used the data from the $Herschel$ PACS 70 and 160\,$\mu$m catalogs instead of IRAS 60 and 100\,$\mu$m data.

\section{Results and discussion}
\label{3}

\subsection{Structure of the molecular cloud}
\label{3.1}
\begin{figure*}
\centering
\includegraphics[width=0.46\linewidth]{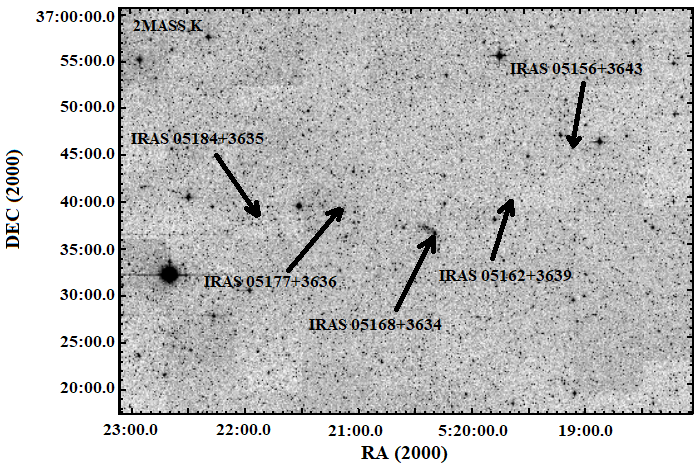}
\includegraphics[width=0.46\linewidth]{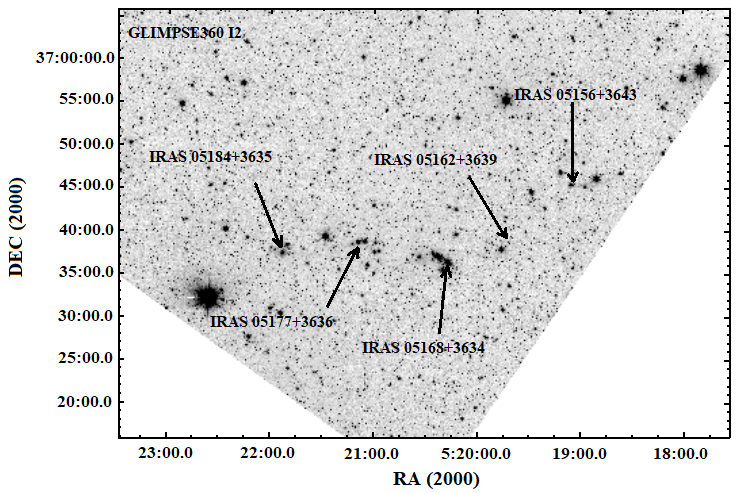}
\includegraphics[width=0.46\linewidth]{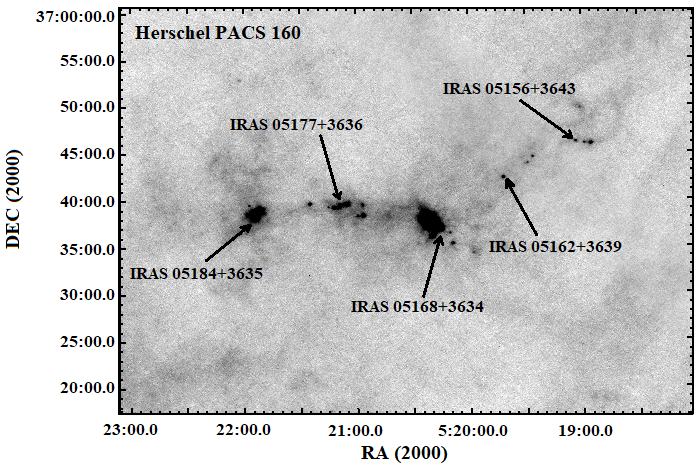}
\includegraphics[width=0.46\linewidth]{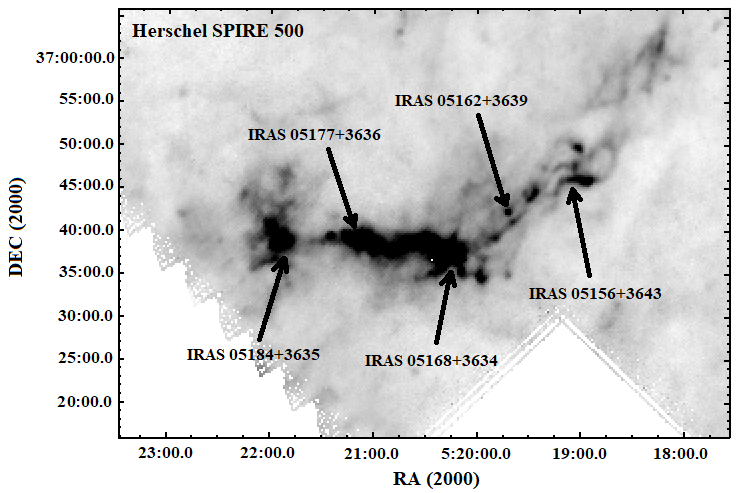}
\caption{Investigated region in different wavelengths. Top left and right panels show the region in NIR (2MASS K band) and MIR (GLIMPSE\,360 I2 band) wavelengths, respectively. Bottom left and right panels show the region in FIR wavelengths: $Herschel$: PACS\,160 and SPIRE\,500, respectively. The position of five IRAS sources are indicated by arrows.}
\label{fig:1}
\end{figure*}

The IRAS\,05168+3634 star-forming region has a more complicated structure in the FIR wavelengths than in the NIR and MIR. The complex structure of the region is clearly visible especially in the images of the $Herschel$ PACS 160\,$\mu$m and SPIRE 250, 350, 500\,$\mu$m bands. Figure \ref{fig:1} shows the region in different wavelengths from NIR to FIR. Moving toward longer wavelengths in the $Herschel$ PACS 160\,$\mu$m and the SPIRE 250, 350, 500\,$\mu$m bands images, the cloud filaments surrounding IRAS\,05168+3634 become more visible and it is obvious that the IRAS\,05168+3634 star-forming region is not limited in 1.5\,arcmin radius from geometric center \citep{azatyan16} but is more extended and is located within a 24\,arcmin radius molecular cloud where the center of the molecular cloud is the conditionally selected source IRAS\,05168+3634. Studying the common star-forming region in the molecular cloud, it turns out that apart from IRAS\,05168+3634, there are four IRAS sources (IRAS\,05184+3635, IRAS\,05177+3636, IRAS\,05162+3639, and IRAS\,05156+3643) embedded in the same molecular cloud. On the MIR [4.5]\,$\mu$m band image can be clearly seen that these IRAS objects, like IRAS\,05168+3634, have a concentration of sources around them. There is very little information about these IRAS sources. IRAS\,05184+3635 and IRAS\,05177+3636 are associated with dark clouds DOBASHI 4334 and 4326, respectively \citep{dobashi11}. In \citet{casoli86} the distances of IRAS\,05184+3635 and IRAS\,05177+3636 were assessed based on the $^{13}$CO velocities: -17\,km/s and -15\,km/s, respectively, as a result of which both IRAS\,05184+3635 and IRAS\,05177+3636  were evaluated at the same 1.4\,kpc distance. The latter value coincides with the distance of IRAS\,05168+3634 based on trigonometric parallax. This also indicates that these IRAS sources are most likely to be found in the same molecular cloud. On the other hand, \citet{wouterloot89}  estimated the distance of IRAS\,05184+3635 based on the CO ($J=1-0$) line parameters, taking into account the same -17\,km/s velocity value, at a distance of 9.4\,kpc. It should be noted that it is hard to say which of the estimated distances is correct. There are two objects near IRAS\,05177+3636 detected at submillimeter wavelengths \citep{di-francesco08}. IRAS\,05162+3639 is associated with the H$_{2}$O maser \citep{sunada07}. A high proper-motion star has been detected in the LSPM-NORTH catalog 0.35\,arcmin from the IRAS\,05156+3643 \citep{lepine05} probably compatible with IRAS\,05156+3643 within the error bars.
         
\begin{figure*}
\centering
\includegraphics[width=0.7\linewidth]{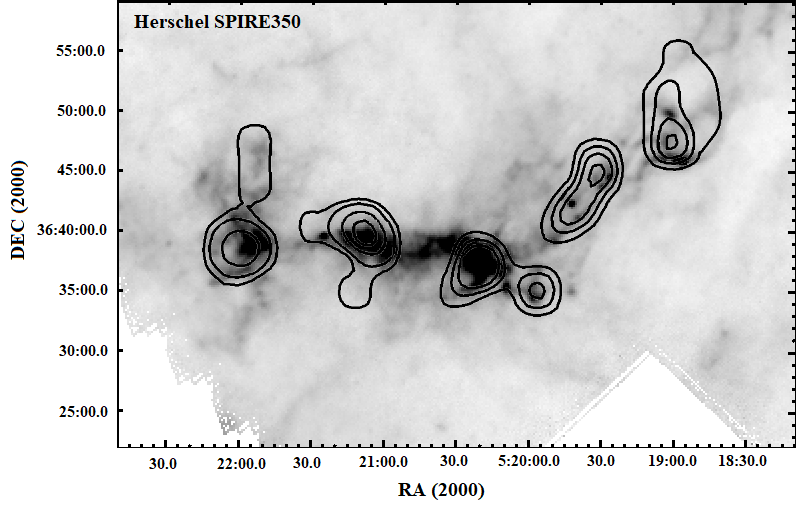}
\caption{Stellar surface density distribution based on $Herschel$ PACS\,160\,$\mu$m photometry. Stellar surface density distribution map is overplotted on $Herschel$ SPIRE 350\,$\mu$m image. The surface density of the last isodences exceeds the average value of the field surface density on 1\,$\sigma$}
\label{fig:2}
\end{figure*}

We have constructed a map of the distribution of stellar surface density within a 48\,\arcmin\,\,$\times$\,48\,\arcmin\, region to investigate the structure of each concentration in the molecular cloud, using photometric data of the $Herschel$ PACS Point Source Catalog: 160\,$\mu$m and Extended Source List (red band). The density was determined simply by dividing the number of stellar objects in a 200\,\arcsec\,\,$\times$\,200\,\arcsec\, area with a step size of 100\,\arcsec\,. Figure \ref{fig:2} shows the map of the distribution of stellar surface density based on $Herschel$ PACS\,160\,$\mu$m photometry.

In the case of the homogeneous distribution of stars in the field, it is known that the number of stars is comparable to the surface area occupied by those stars. We wanted to determine how much the distribution of stars in this field differs from the homogeneous distribution. Table \ref{tab:1} shows the surface area of all subregions according to the isodences in Col. 2 and the number of stars on those surface areas in Col. 3. The labels S$_{TR}$ and S$_{Total}$ are the total surface area of five subregions and the surface area within a 24\,arcmin radius (IRAS\,05168+3634 is considered the center), respectively. The ratios of S$_{TR}$/S$_{Total}$=0.02 and N$_{TR}$/N$_{Total}$=0.7 indicate what fraction of the surface area within a 24\,arcmin radius is the total surface area of the subregions and how much of the objects in the whole field are on the total surface area of the subregion, respectively.
\begin{table}
\caption[]{Sizes and contents of the subregions}
\resizebox{0.48\textwidth}{!}{
\label{tab:1}
\centering
\begin{tabular} {l c c}
\hline  \hline \noalign{\smallskip}

Name &  Surface area  & Objects       \\
 & (arcmin$^2$)    &         \\ \hline \noalign{\smallskip}
(1)             &       (2)     &       (3)     \\
\hline \noalign{\smallskip}
IRAS\,05184+3635  &       5.245   &       6       \\
IRAS\,05177+3636  &       10.072  &       7       \\
IRAS\,05168+3634  &       11.14   &       7       \\
IRAS\,05162+3639  &       5.384   &       6       \\
IRAS\,05156+3643  &       10.068  &       6       \\
&       S$_{TR}$= 41.909     &       N$_{TR}$= 32 \\
&       S$_{Total}$=1808.64  &       N$_{Total}$=43       \\
\hline 
\end{tabular}
}
\tablefoot{
(1) Name of the subregions, (2) Approximate surface area of each subregion based on $Herschel$ PACS\,160\,$\mu$m catalog data, (3) Number of objects in each subregion based on $Herschel$ PACS\,160\,$\mu$m catalog data.}

\end{table}
From these values, it is possible to estimate the difference between the distribution of stars in our field and the homogeneous distribution. As a result, the distribution of stars in our field is 35 times different from the homogeneous distribution, which is a fairly high value. On the other hand, the distribution of sources in the field repeats the shape of the molecular cloud seen in FIR wavelengths, which suggests that the distribution of these subregions is not accidental, that is, they are connected to each other and belong to the same molecular cloud. From this, we can conclude with high probability that all five IRAS star-forming regions are at the same distance. Since only the IRAS\,05168+3634 star-forming region has estimated distances (see Section \ref{1}), we can say that IRAS\,05184+3635, IRAS\,05177+3636, IRAS\,05162+3639, and IRAS\,05156+3643 are also located at a distance of 6.08\,kpc (kinematic estimation) or 1.88$^{+0.21}_{-0.17}$\,kpc (based on the trigonometric parallax). These distance estimations are used in future calculations in this paper.
\subsection{Selection of objects}
 \label{3.2}

For the selection of objects in the molecular cloud, we used the data of NIR, MIR, and FIR catalogs (see Section \ref{2}) within a radius of 24\,arcmin concerning the conditionally selected IRAS\,05168+3634. We chose GPS UKIDSS-DR6 as the main catalog, and the other catalogs were cross-matched with it. As was mentioned in Section \ref{2}, the GPS UKIDSS-DR6 catalog for individual objects provides a probability (in percent) of being a star, galaxy, and noise; therefore, we selected objects with a probability of being noise <\,50\%, and taking into account the completeness limit of UKIDSS survey in K band, the objects that have a measured magnitude of K$\geqslant$18.02 were removed from the list. This yielded approximately 48000 objects. The MIR and FIR photometric catalogs were cross-matched with the GPS UKIDSS-DR6 catalog within 3\,$\sigma$ of combined error matching radius (Col. 3 in Table \ref{tab:2}). Matching radii are evaluated taking into account the positional accuracy of each catalog (Col. 2 in Table \ref{tab:2}). However, several catalogs provide photometric measurements for extended sources, for which positional errors are larger than values provided in catalogs. These catalogs are the $Herschel$ PACS Point Source Catalog: Extended Source List and the $Herschel$ SPIRE Point Source Catalog: 250,350,500\,$\mu$m; therefore, the cross-match was done by eye for these catalogs and then a combined photometric catalog was obtained. In the next sections, the set of steps that we used to identify the YSOs.

\begin{table}
\caption[]{Properties of used catalogs}
\resizebox{0.48\textwidth}{!}{
\label{tab:2}
\begin{tabular}{l c c l}
\hline\hline\noalign{\smallskip}
\centering
Catalog name    &       Positional accuracy     &       3\,$\sigma$ of combined error    &       Reference       \\
        &       (arcsec)       &       (arcsec)       &  \\
\hline\noalign{\smallskip}
(1)     &       (2)     &       (3)     &       (4)     \\
\hline \noalign{\smallskip}
GPS UKIDSS-DR6  &       0.3     &       $-$     &       1 \\
GLIMPSE\,360      &       0.3     &       1.2     &       2    \\
ALLWISE &       1       &       3       &       3    \\
MSX     &       3.3     &       10      &       4     \\
IRAS    &       16      &       48      &       5  \\
PACS:Extended source    &       2.4     &    7.2     &   6 \\
PACS: 70\,$\mu$m     &       1.5     &       5       &       7   \\
PACS: 160\,$\mu$m    &       1.7     &       5.2     &      8   \\
SPIRE:250,350,500\,$\mu$m    &   1.7   &       5.2     &       9    \\
\hline
\end{tabular}

}
\tablebib{
1. \citet{lucas08}; 2. \citet{churchwell09}; 3. \citet{wright10}; 4. \citet{egan03};
5. \citet{neugebauer84}; 6. \citet{maddox17}; 7. \citet{poglitsch10}; 8. \citet{poglitsch10};
9. \citet{griffin10}.
}
\tablefoot{
(1) Name of  catalog used, (2) Positional accuracy of each catalog, (3) 3\,$\sigma$ of combined error of cross-matched catalogs, (4) Source of data used.}
\end{table}


\subsection{Color-color diagrams}
\label{3.3}
According to the star formation theory, the IR excess of young stars is caused by a circumstellar disk and gas-dust envelope, which are known as two of the main characteristics of YSOs \citep{lada03, hartmann09}. Therefore, it is possible to carry out the selection of YSOs in the region as they contain the IR excess in the NIR and/or MIR ranges.

One of the most powerful tools for identifying YSO candidates via reddening and excess is their location on color-color diagrams. The choice of colors depends on the available data. We have constructed four color-color diagrams. The investigated region is quite large and there is a high probability of selecting objects that do not belong to the molecular cloud as a result of photometric measurement errors; therefore, we chose as YSOs those stars that are classified as objects with IR excess in at least two color-color diagrams to minimize the likelihood of choosing incorrectly.

\begin{figure*}
\centering
\includegraphics[width=0.46\linewidth]{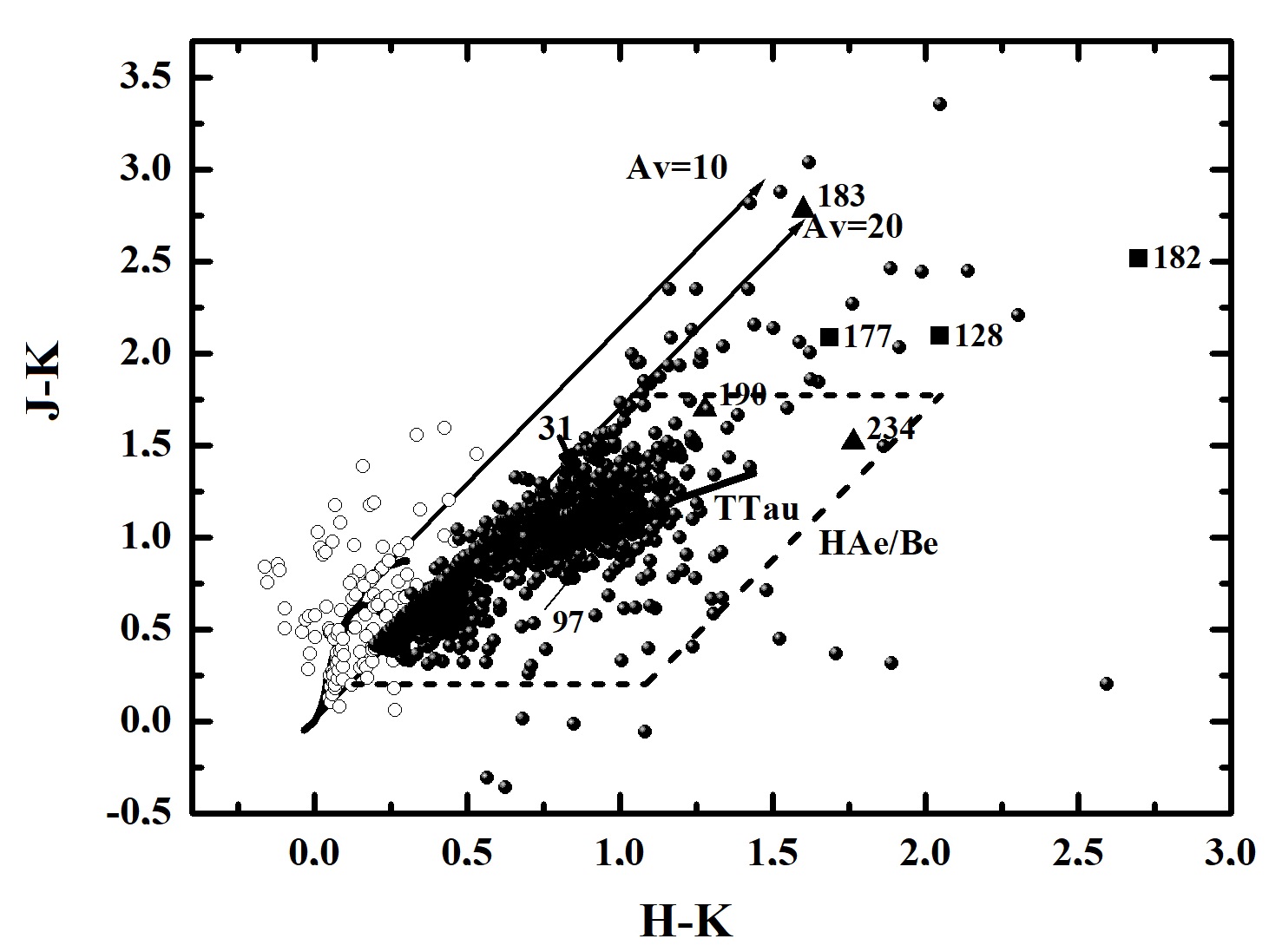}
\includegraphics[width=0.46\linewidth]{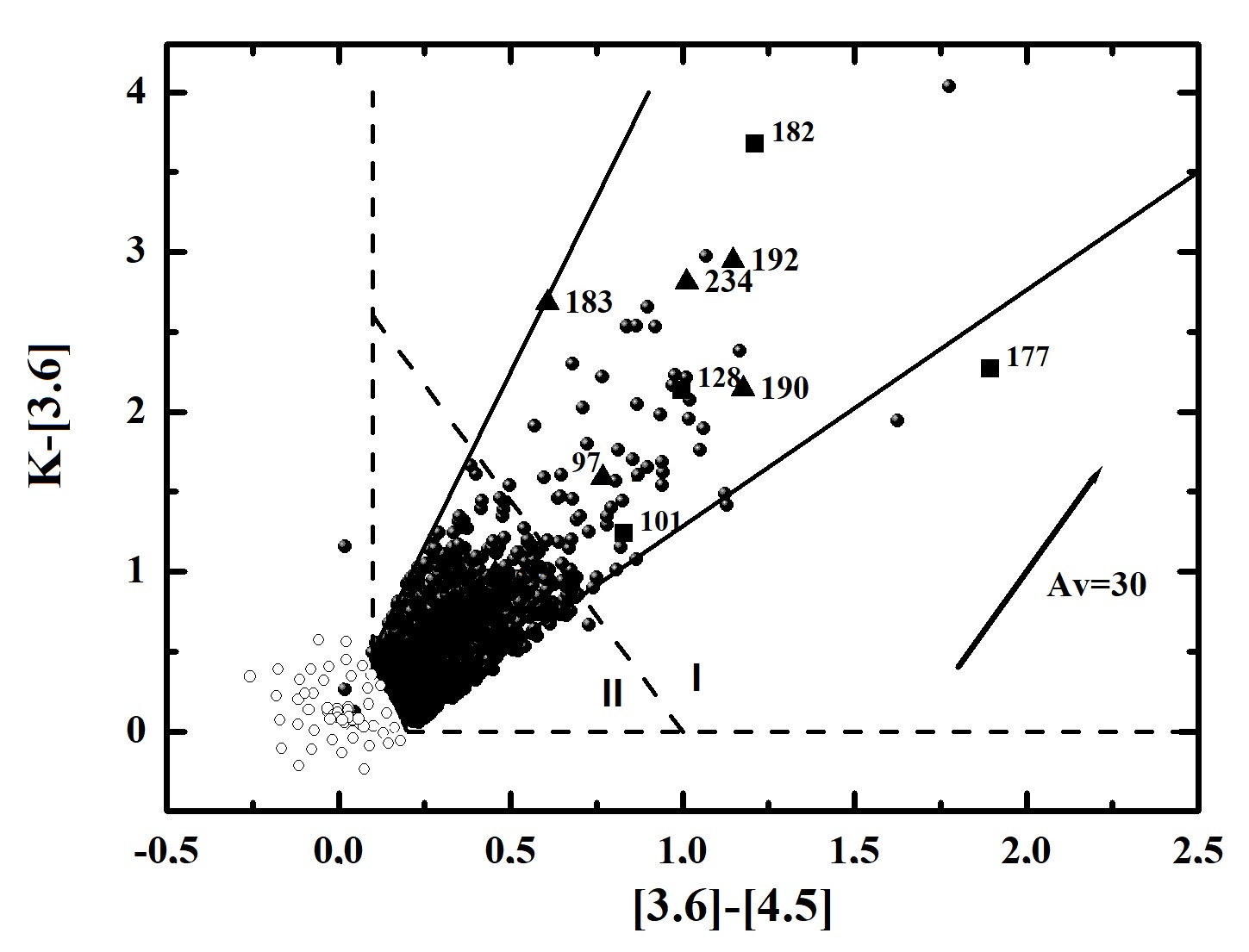}
\includegraphics[width=0.45\linewidth]{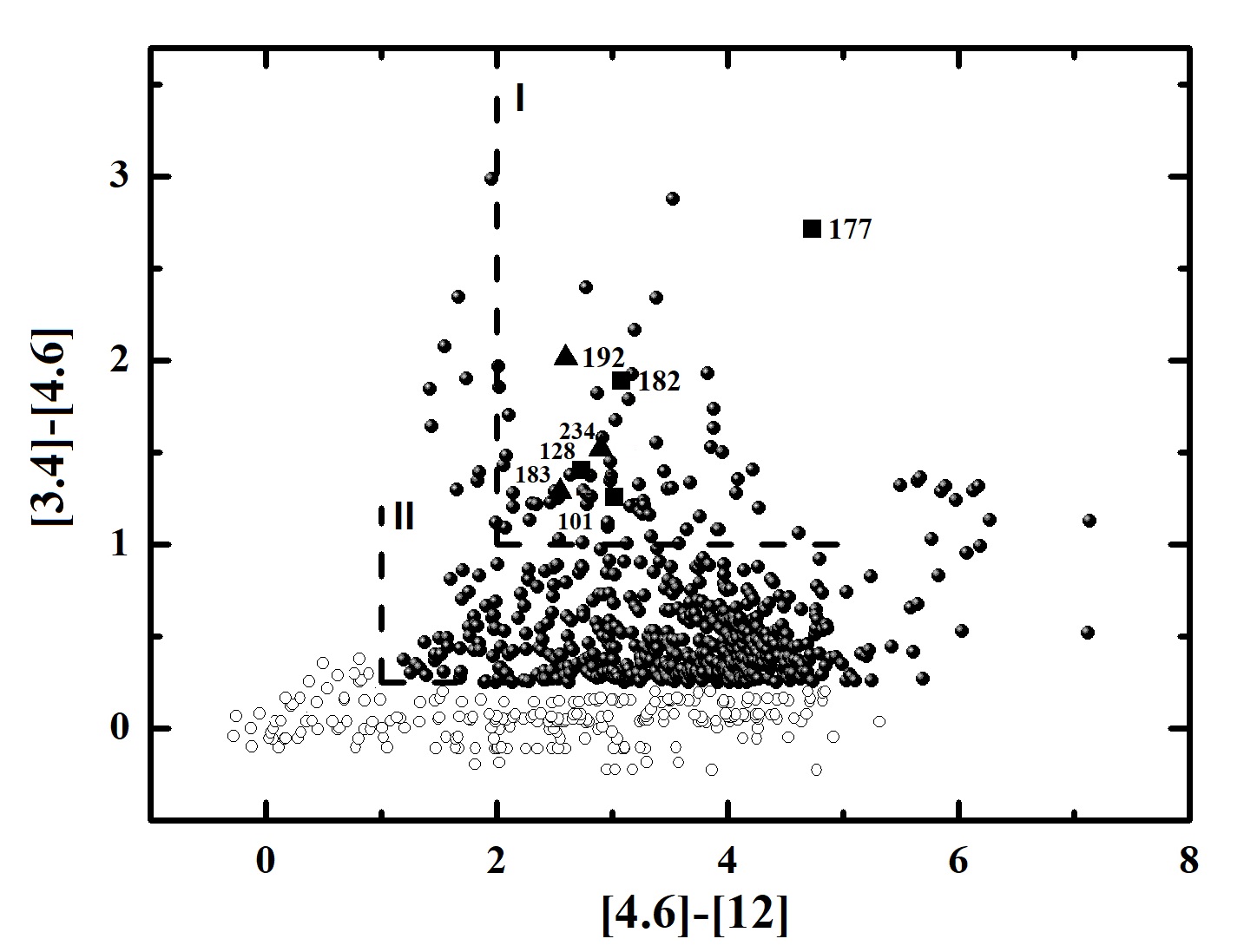}
\includegraphics[width=0.46\linewidth]{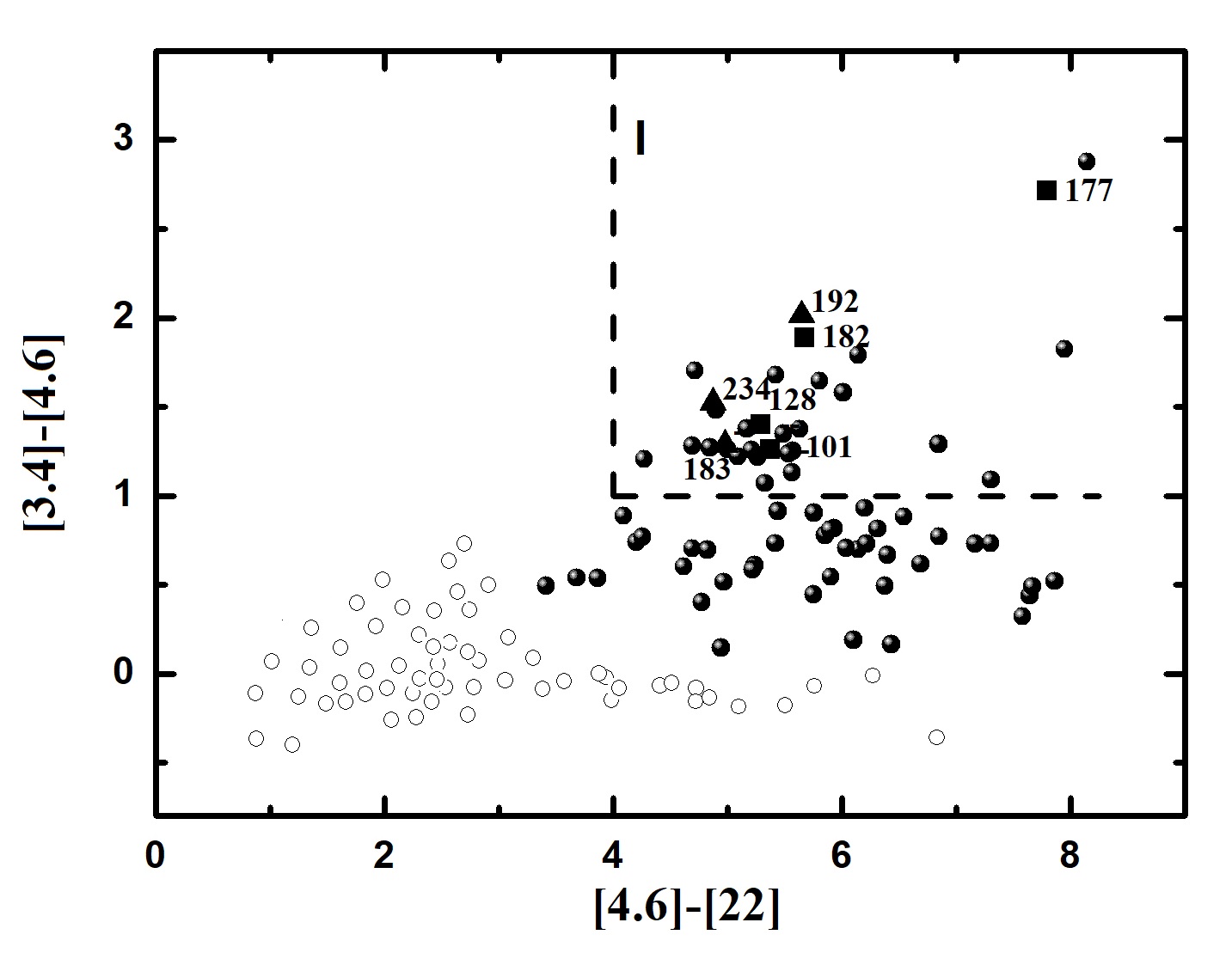}
\caption{Color-color diagrams of the region. In the top left panel is shown the (J-H) vs. (H-K) diagram. The dwarf and giant loci (solid and dashed curves, respectively) are  from \citet{bessell88} and were converted to the CIT system \citep{carpenter01}. The parallel lines represent the interstellar reddening vectors \citep{rieke85}. The locus of unreddened classical T Tauri stars is from \citet{meyer97}, and the region bounded by dashed lines is the Herbig\,Ae/Be stars location \citep{hernandez05}. In the top right panel the K-[3.6] vs. [3.6]-[4.5] diagram is presented. In this diagram Class\,I and II domains are separated by the dashed line. The arrow shows the extinction vector \citep{flaherty07}. All the lines in the K-[3.6] vs. [3.6]-[4.5] diagram are from \citet{allen07}. In bottom left and right panels are shown the [3.4]-[4.6] vs. [4.6]-[12] and [3.4]-[4.6] vs. [4.6]-[22] diagrams. The filled circles are selected YSOs and open circles are unclassified ones. Not all unclassified objects are presented in these diagrams. IRAS and MSX sources are indicated by triangles and squares, respectively, and they are labeled (see Tables \ref{tab:5} and \ref{tab:6}).}
\label{fig:3}
\end{figure*}

The first IR excess object identification was carried out with (J-H) versus  (H-K) color-color diagram. Figure \ref{fig:3} (top left panel) shows the (J-H) versus (H-K) color-color diagram where the solid and dashed curves represent the locus of the intrinsic colors of dwarf and giant stars, taken from \citet{bessell88} after being converted to the CIT system using the relations given by \citet{carpenter01}. The parallel solid lines drawn from the base and tip of the dwarf and giant loci, are the interstellar reddening vectors \citep{rieke85}. The locus of unreddened classical T Tauri stars (CTTSs) is taken from \citet{meyer97}. The region bounded by the dashed lines where the pre-main sequence (PMS) stars with intermediate mass, i.e., Herbig\,Ae/Be stars are usually found \citep{hernandez05}. The objects with different evolutionary stages have certain places in this diagram \citep{lada92}: Classical Be stars, objects located to the left and objects located to the right of reddening vectors. Classical Be stars have J-K < 0.6 and H-K < 0.3 color indexes, and we removed them from the list. There is no real physical explanation for those objects which are located to the left of reddening vectors, so we also removed these objects. However, it is possible that among Classical Be stars and objects located to the left of reddening vectors that there will be objects that belong to a molecular cloud. For further study, we chose objects located to the right of reddening vectors and the deviation of YSOs from the main sequence (MS) on this diagram can have two causes: the presence of IR excess and interstellar absorption, which also leads to the reddening of an object. In the latter case, the deviation from MS will be directed along reddening vectors. We have selected as YSOs those stars that have a considerable, accurately expressed IR excess.

The reddening of classified YSOs cannot be caused only by interstellar absorption and, at least partially, IR excess is created by presence of a circumstellar disk and an envelope. Therefore, the objects that are located to the right of reddening vectors in the (J-H) versus (H-K) diagram can be considered YSO candidates. Among the selected YSOs and objects within the reddening band of the MS and giant, we classified as Class\,I evolutionary stage objects those which have (J-K) > 3 mag color index \citep{lada92} and located in the top right in the (J-H) versus (H-K) diagram. Other objects in the reddening band are generally considered to be either field stars or Class\,III objects with comparably small NIR excess, and making the right selection among them is very difficult, so we added to our list those objects classified as YSOs in at least two other color-color diagrams.

Not all objects in the main sample were detected in J, H, K bands simultaneously; therefore, we  used the data of the GLIMPSE\,360 catalog to construct color-color diagram combining NIR and MIR photometry in order to identify sources with IR excesses and to compile a more complete excess/disk census for the region. Since the 4.5\,$\mu$m band is the most sensitive band to YSOs of the four IRAC bands \citep{gutermuth06}, we used the K-[3.6] versus [3.6]-[4.5] color-color diagram. Figure \ref{fig:3} (top right panel) shows the K-[3.6] versus [3.6]-[4.5] color-color diagram, where the diagonal lines outline the region of location of YSOs with both Class\,I and Class\,II evolutionary stages. The de-reddened colors are separated into Class\,I and II domains by the dashed line. Arrow shows the extinction vector \citep{flaherty07}. All the lines in the K- [3.6] versus [3.6]-[4.5] diagram are taken from \citet{allen07}. Both the (J-H) versus (H-K) and the K- [3.6] versus [3.6]-[4.5] color-color diagrams are widely used for the separation of YSOs with different evolutionary classes \citep{meyer97,gutermuth06,allen07}.

We also constructed two other color-color diagrams using the list of objects with good WISE detections, i.e., those possessing photometric uncertainty <\,0.2\,mag in WISE bands. Figure \ref{fig:3} (lower left panel) shows the [3.4]-[4.6] versus [4.6]-[12] color-color diagram. As  was mentioned for the previous color-color diagrams, the objects with different evolutionary stages are located in certain places in this diagram too \citep{koenig12}, i.e., Class\,I YSOs are the reddest objects, and are selected if their colors match [3.4]-[4.6] > 1.0 and [4.6]-[12] > 2.0. Class\,II YSOs are slightly less red objects and are selected with colors [3.4]-[4.6]-$\sigma$([3.4]-[4.6]) > 0.25 and [4.6]-[12]-$\sigma$([4.6]-[12]) > 1.0, where $\sigma$(...) indicates a combined photometric error, added in quadrature.

We can check the accuracy of previous classification of stars that possess photometric errors <\,0.2\,mag in WISE band 4. Previously classified Class\,I sources were re-classified as Class\,II if [4.6]-[22] < 4.0 and Class\,II stars have been placed back in the unclassified pool if [3.4]-[12] < -1.7\,$\times$\,([12]-[22]) + 4.3 \citep{koenig12}. However, there are no incorrect selections in the pre-classified objects with 1, 2, 3 bands and it confirms the results obtained in the [3.4]-[4.6] versus [4.6]-[12] color-color diagram (Figure \ref{fig:3}, lower right panel).

In total, we selected 1224 YSOs within a 24\,arcmin radius, and they are indicated with filled circles. IRAS and MSX sources are indicated by triangles and squares, respectively, and they are labeled in the diagrams. Unfortunately, with this kind of selection of YSOs we cannot find the exact number of YSOs. T Tauri objects with comparably small amounts of NIR excesses may be located in the reddening band, and therefore are excluded from the selection. On the other hand, there can be fore-/background objects among the selected YSOs.

\begin{figure*}
\centering
\includegraphics[width=0.6\linewidth]{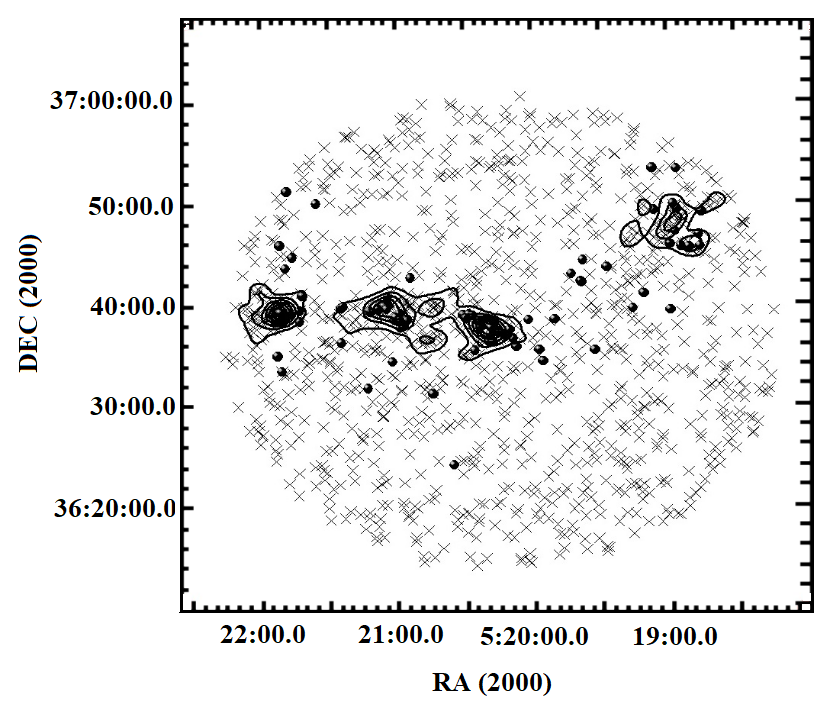}
\caption{Stellar surface density distribution based on color-color diagrams. Class\,I and Class\,II evolutionary stage objects are indicated by filled circles and crosses, respectively.}
\label{fig:4}
\end{figure*}

Figure \ref{fig:4} shows the distribution of classified YSOs in the field. Class\,I and Class\,II objects are indicated by filled circles and crosses, respectively. It can be clearly seen from the image that Class\,II objects are distributed more homogeneously on the field than Class\,I objects, which are located in certain areas and show clear concentrations. This confirms the assumption that, unlike the Class\,II objects, Class\,I objects did not have enough time to leave their birthplaces after formation. There are several Class\,I objects in the field that are far from the main concentrations, and because the distribution of Class\,I objects is closer to the initial distribution it can be argued that the Class\,I objects either have some measurement errors in magnitudes or that they do not belong to the molecular cloud. Since the distribution of stars and clearly visible concentrations repeat the shape of the molecular cloud, the probability of superposition is very small because in that case there would be other similar concentrations in the field, which are not detected outside the molecular cloud filaments.
 
Since the region is quite large, the probability of being fore-/background objects within the selected Class\,II objects is very high, and in the case of Class\,I objects that possibility is small as they are not detected outside the molecular cloud filaments. In order to minimize the number of incorrect Class\,II objects, further investigations will only be performed on concentration areas. For that purpose, we have constructed map of the distribution of stellar surface density within a 48\,\arcmin\,\,$\times$\,48\,\arcmin\, region to investigate the structure and size of each concentration in the molecular cloud, using the coordinates of selected YSOs. The contours of the distribution of stellar surface density is overplotted on Figure \ref{fig:4} and four subregions are clearly seen, each one surrounding one IRAS source. We refined the radius of each subregion relative to their geometric centers based on the density distribution of selected YSOs. The stellar density was determined for each ring of width 0.1\,\arcmin\, by dividing the number of stars by the surface area. The radius of the subregions was considered the value from which according to Poisson statistics the fluctuations of the stellar density in the rings become random. Table \ref{tab:3} presents the coordinates of IRAS sources in Cols. 2 and 3, the coordinates of geometric centers are in Cols. 4 and 5, and  the radii based on stellar density distribution are in the last column. There is no evidence of a real concentration only around IRAS\,05162+3639. On the other hand, two objects from the GPS UKIDSS-DR6 catalog were identified with IRAS\,05162+3639 as a result of a cross-match, i.e.,  IRAS\,05162+3639 is probably a binary object and three more Class\,I objects are located very close to it, but they do not show any real concentration. Therefore, the 0.25\,arcmin value of radius given in Table \ref{tab:3} is conditional and includes these three Class\,I objects and the binary associated with IRAS\,05162+3639. As was already mentioned above, the objects within the determined radii will be explored in greater detail, so further studies will be conducted for the 240 YSOs of the 1224 selected from color-color diagrams; this total number of objects falls within the already-defined radii of five subregions. However, with this kind of choice, we lose a large number of objects that are actually located in the molecular cloud filaments but are left out of the determined radius of subregions. Otherwise, we would increase the number of non-members. Therefore, in this case, it would be possible to insist that the objects belonging to the molecular cloud were chosen as accurately as possible and that each subregion represents a separate star-forming region. Table \ref{tab:5} shows the catalog of 240 selected YSOs with NIR and MIR photometry, while Table \ref{tab:6} presents the fluxes for those YSOs which have FIR photometry. In the $Herschel$ SPIRE 250, 350, 500\,$\mu$m catalogs, the photometry of objects was determined by four methods: Timeline Fitter value (TML), Daophot, Sussextractor, and Timeline Fitter 2 (TM2). For the slightly extended sources that were accepted in the $Herschel$ SPIRE 250, 350, 500\,$\mu$m catalog, the TM2 value provides the best guess for an extended flux \citep{griffin10}. All selected YSOs in subregions that have measured fluxes at the 250, 350, 500\,$\mu$m wavelengths are classified as extended sources in the $Herschel$ SPIRE 250, 350, 500\,$\mu$m catalogs. Taking into account the above, only TM2 photometry is presented in Table \ref{tab:6}.

\begin{table}
\caption[]{Geometric centers of subregions}
\resizebox{0.49\textwidth}{!}{
\label{tab:3}
\begin{tabular}{l c c c c c}
\hline\hline
\noalign{\smallskip}
\centering
IRAS & $\alpha$(2000) & $\delta$(2000) & $\alpha$(2000) & $\delta$(2000) & Radius \\
 & (hh mm ss) & (dd mm ss) & (hh mm ss) & (dd mm ss) & (arcmin) \\
\hline\noalign{\smallskip}
(1) & (2) &  (3) & (4) & (5) & (6) \\
\hline \noalign{\smallskip}
05184+3635 & 05 21 53.2 & +36 38 20.4 & 05 21 52.6 & +36 39 07.1 & 2.5 \\
05177+3636 & 05 21 09.4 & +36 39 37.1 & 05 21 02.8 & +36 38 28.5 & 3.5 \\
05168+3634 & 05 20 16.4 & +36 37 18.7 & 05 20 22.3 & +36 37 33.9 & 3 \\
05162+3639 & 05 19 38.4 & +36 42 25.0  & 05 19 38.4 & +36 42 25.0 & 0.25 \\
05156+3643 & 05 19 03.6 & +36 46 15.7 & 05 19 04.0 & +36 48 02.0 & 2.8 \\
\hline\noalign{\smallskip}
\end{tabular}
}
\tablefoot{
(1)-Name of subregions, (2),(3)-The coordinates of IRAS sources, (4),(5)-The coordinates of geometric centers, (6)-The radius of each subregion according to YSOs surface density distribution in the molecular cloud}
\end{table}


\subsection{Color-magnitude diagrams}
\label{3.4}
\begin{figure*}
\centering
\includegraphics[width=0.49\linewidth]{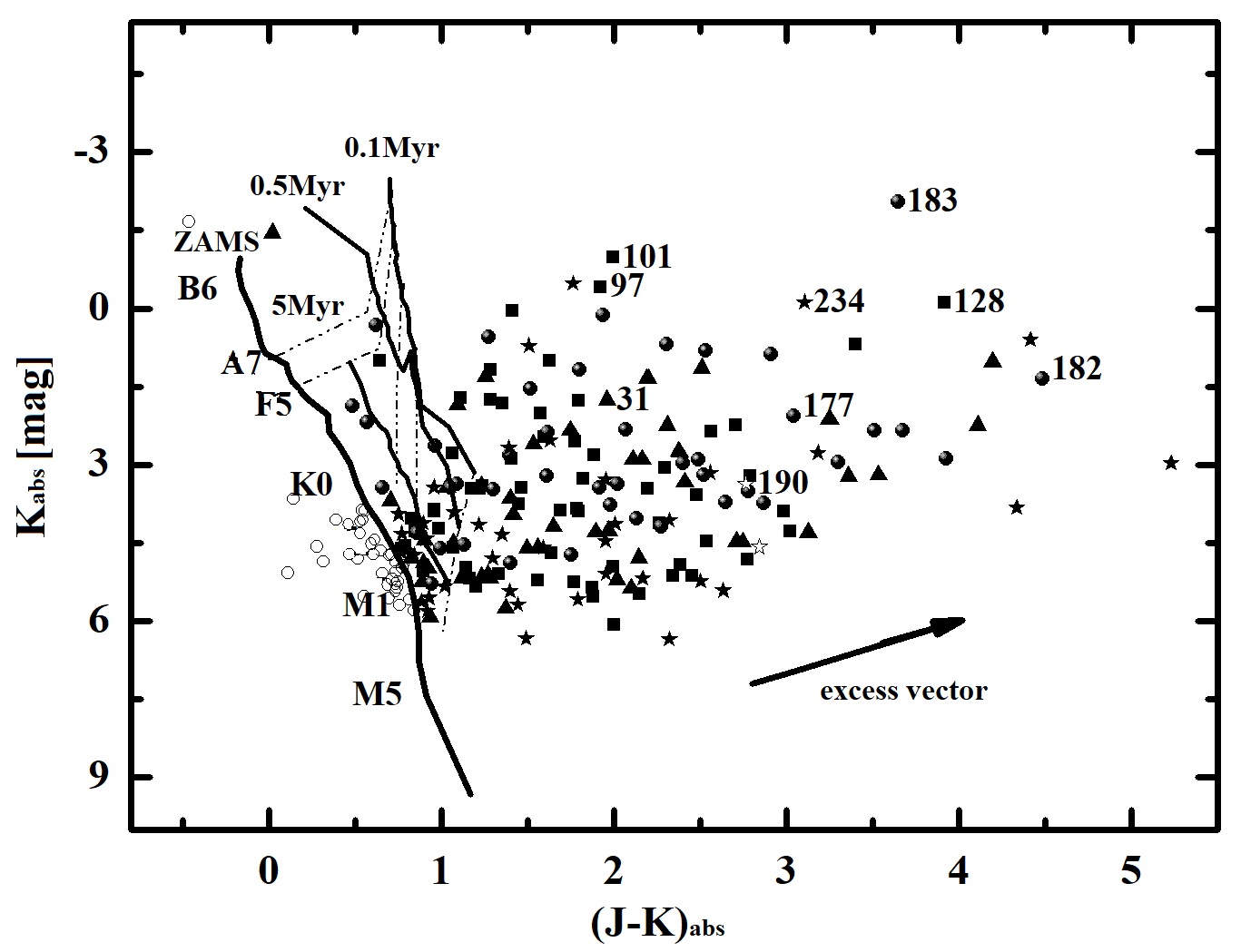}
\includegraphics[width=0.49\linewidth]{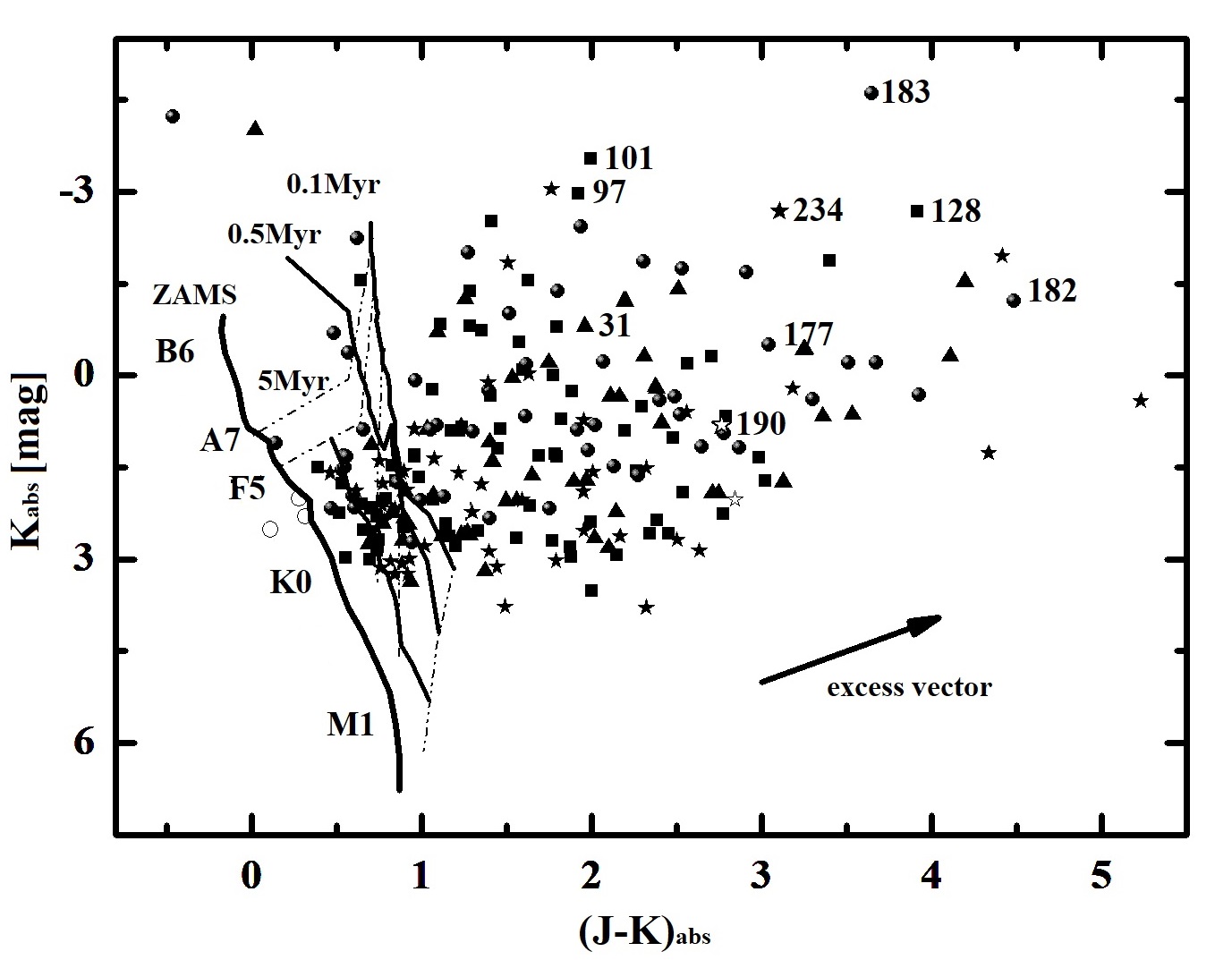}
\caption{K vs. (J-K) color-magnitude diagrams for the identified YSOs in subregions. The PMS isochrones for 0.1, 0.5, and 5\,Myr by \citet{siess00} and ZAMS are drawn as solid thin and thick lines, respectively. Dashed lines represent the evolutionary tracks for different masses. The positions of a few spectral types are labeled. The J and K magnitudes of the YSOs are corrected for the interstellar extinctions determined according to the COBE/DIRBE and IRAS/ISSA maps and distances 1.88\,kpc (left panel) and 6.1\,kpc (right panel), respectively. The arrow indicates the average slope of NIR excesses caused by disks around YSOs, as determined by \citet{lopez07}. The objects belonging to different subregions are shown as follows: IRAS\,05184+3635 (triangles), IRAS\,05177+3636 (square), IRAS\,05168+3634 (filled circles), IRAS\,05162+3639 (open stars), and IRAS\,05156+3643 (filled stars). Open circles  indicate  objects  considered  members of  subregions, but in the color-magnitude diagram they are located to the left of ZAMS. The IRAS and MSX sources are labeled (see Table \ref{tab:5}).}
\label{fig:5}
\end{figure*}

The color-magnitude diagram is a useful tool for studying the nature of the stellar population within star-forming regions and for estimating its spectral types. The distribution of the 240 identified YSOs in the K versus J-K color-magnitude diagrams are shown in Figure \ref{fig:5} with different symbols for each subregion. We use the K versus J-K diagram because with this combination we get the maximum contrast between the IR excess produced by the presence of the disk, which affects mostly K, and the interstellar extinction, which has a greater effect on J. In Figure \ref{fig:5} the zero age main sequence (ZAMS, thick solid curve) and PMS isochrones for ages 0.1, 0.5, and 5\,Myr (thin solid curves) are taken from \citet{siess00}. We used the conversion table of \citet{kenyon94}. The J and K photometry of the selected YSOs are corrected for two different distances: 6.1 and 1.88\,kpc because, as mentioned above, all subregions with high probability are located at the same distance (see Section \ref{3.1}).

According to the COBE/DIRBE and IRAS/ISSA maps \citep{schlegel98}, we have estimated the interstellar extinction (A$_v$) values toward five IRAS sources provided in Table \ref{tab:4}. There is data of observations of $^{13}$CO cores, providing N$_{H_{2}}=4.5\,\times\,10^{21}\,cm^{-2}$ column density \citep{guan08}, only in the IRAS\,05168+3634 star-forming region and using the conversion factor between column density and interstellar extinction N$_{H_{2}}=9.4\,\times\,10^{20}\,cm^{-2}$ (A$_v$\,mag) \citep{bohlin78} have we  received a value of A$_v$=4.8\,mag for interstellar extinction. The average of both interstellar extinction values (A$_v$=4.5\,mag) for the IRAS\,05168+3634 star-forming region is used to correct the J and K magnitudes. Correction of the J and K magnitudes for the other four regions was done with the interstellar extinction estimated according to the COBE/DIRBE and IRAS/ISSA maps.

In general, the selected YSOs (according to the color-color diagrams) are distributed to the right of 0.1\,Myr isochrones especially in the case of 6.1\,kpc distance, and this distribution confirms that they are YSOs. The YSOs associated with IRAS and MSX sources are located within the range of the largest IR excess sources in both color-magnitude diagrams. Only a few identified as YSOs objects in the color-color diagrams are located to the left (open circles) of ZAMS, which means non-membership, i.e., they are probably fore-/background objects. At different distances, the numbers of non-members are different, but it must be taken into account that the J and K magnitude corrections we made for objects belonging to an individual subregion using interstellar extinction values estimated toward IRAS sources only, which means those values would be different from the interstellar extinction values toward each object, and as a result some objects fall left of ZAMS. Thus, those objects were not removed from the list of YSOs (Table \ref{tab:5}), but are shown as open circles in both color-magnitude diagrams. The results of the selection with the color-color and color-magnitude diagrams are presented in the Table \ref{tab:4} separately for each subregion depending on distance.

According to the results of color-color and color-magnitude diagrams, within the selected radius of each subregion the youngest is the IRAS\,05168+3634 star-forming region since Class\,I objects represent a fairly large percentage. The next youngest subregion by Class\,I objects is IRAS\,05177+3636, and it should be noted, even though there is no real concentration around  IRAS\,05162+3639, that only the Class\,I objects are inside the conditional selected radius. Finally, the subregions around IRAS\,05184+3635 and IRAS\,05156+3643 in the outer part of the molecular cloud are the oldest. However, it must be kept in mind that Class\,II objects have had enough time after formation and have left their birthplaces, so perhaps most of them are beyond the determined dimensions of subregions. Table \ref{tab:4} shows the percentage of Class\,I objects in each subregion.

The distribution of the YSOs on the K versus J-K color-magnitude diagram can be used to estimate their approximate age and mass based on the evolutionary models for objects with ages older than 1\,Myr \citep{preibisch12}. It is evident from Figure \ref{fig:5} that the YSOs in the molecular cloud in general are distributed to the right of 1\,Myr isochrone, where the estimation of age and mass of those YSOs will be incorrect.

However, the color-magnitude diagrams show a certain distribution of mass in each subregion. Let us try to estimate the mass limit for each subregion. Since NIR color excess, displayed in the color-color diagrams (Figure \ref{fig:3}), is usually caused by the presence of disks around young stars, then by incorporating theoretical accreting disk models, the excess effect on the K versus J-K diagram can be accurately represented by vectors of approximately constant slope for disks around Class\,II T Tauri stars. The components of the vector are (1.01, -1.105) and (1.676, 1.1613) in magnitude units \citep{lopez07}. More massive YSOs are usually much more embedded than T Tauri stars, and that correction is unlikely to apply to such objects. However, the presence of a spherical envelope around the disk should cause a greater decrease in J-K for the same variation in K, than in the case of a ``naked'' disk \citep{cesaroni15}. Therefore, the \citet{lopez07} correction can be used to obtain a lower and an upper limit to the mass of each subregion for two different distances. Table \ref{tab:4} shows the results of mass completeness limit determination. Objects located in the IRAS\,05184+3635, IRAS\,5168+3634, and IRAS\,05156+3643 star-forming regions are explicitly targeted at specific mass ranges for two distances, and for objects in the IRAS\,05177+3636 star-forming region there is no similar dependency. It can be seen in color-magnitude diagrams and from estimated values of mass completeness limit shown in Table \ref{tab:4} Cols. 10 and 11. For the total content of four real subregions based on excess vector analysis, the following result is derived. At a distance of 1.88\,kpc, approximately 80\,\% of the total content of the subregions have < 1 solar mass, and the remaining $\backsim$20\,\% objects have 1--3 solar masses. Only two objects have > 7 solar masses that are clearly distinguished from the rest of the objects in the color-magnitude diagrams. At a distance of 6.1\,kpc, approximately 20\,\% of the total content of the subregions have < 1 solar mass, about 70\,\% objects have 1--3 solar masses,  and  the mass was estimated in the range 3--7 solar masses  in only 7\,\% of the cases. There are only five objects with mass > 7 solar masses.

\begin{table*}
\caption[]{Characteristics of subregions}
\resizebox{1\textwidth}{!}{
\label{tab:4}
\begin{tabular}{*{12}{c}}
\hline \hline
\noalign{\smallskip}
\centering
IRAS    &       \multicolumn{2}{c}{CCD}             &       \multicolumn{2}{c}{CMD-1.88\,kpc}          &       \multicolumn{2}{c}{CMD-6.1\,kpc}     &       Class\,I       &       A$_v$      &       \multicolumn{2}{c}{Mass limit}            &       $\alpha$ slope                   \\
        &       Mem.    &       Class\,I &       Mem.    &       Class\,I &       Mem.    &       Class\,I       &               &       (mag)   &       \multicolumn{2}{c}{M$_{sun}$}     &                               \\
\noalign{\smallskip}
        &               &       (\%)     &               &       (\%)     &               &       (\%)     &               &               &       1.88kpc &       6.1kpc  &                               \\
\hline \noalign{\smallskip}
(1)     &       (2)     &       (3)     &       (4)     &       (5)     &       (6)     &       (7)     &       (8)     &       (9)     &       (10)    &       (11)    &               (12)            \\
\noalign{\smallskip}
\hline \noalign{\smallskip}
05184+3635      &       52      &       21      &       48      &       23      &       52      &       21      &       11      &       1.4     &       0.3-1.5 &       0.8-5   &       0.12    $\pm$   0.04    \\
05177+3636      &       79      &       28      &       65      &       34      &       79      &       28      &       22      &       1.34    &       0.2-2.2 &       0.7- >7  &       0,2     $\pm$   0.02    \\
05168+3634      &       57      &       43      &       45      &       54      &       56      &       45      &       24      &       4.3 (4.5)   &       0.5-2.5 &       0.9- >7  &       0.21    $\pm$   0.05    \\
05162+3639      &       5       &       $-$     &       5       &       $-$     &       5       &       $-$     &       5       &       1.23    &       $-$     &       $-$     &       $-$                     \\
05156+3643      &       47      &       20      &       40      &       23      &       47      &       20      &       9       &       1.03    &       0.2-1.6 &       0.6-3   &       0.15    $\pm$   0.04    \\
\noalign{\smallskip} \hline
\end{tabular}
}
\tablefoot{
(1) Name of subregions, (2) and (3) Number of YSOs and the fraction of Class\,I objects (in percent) according to the color-color diagrams, (4)--(7) Number of YSOs and the fraction of Class\,I objects (in percent) according to the color-magnitude diagrams in different distances, (8) Number of Class\,I evolutionary stage objects, (9) Interstellar extinction according to the COBE/DIRBE and IRAS/ISSA maps, (10)--(11) Mass completeness limits of each subregion in two distances, (12) The $\alpha$ slope of KLF of each subregion}

\end{table*}
\begin{figure*}
\centering
\includegraphics[width=0.484\linewidth]{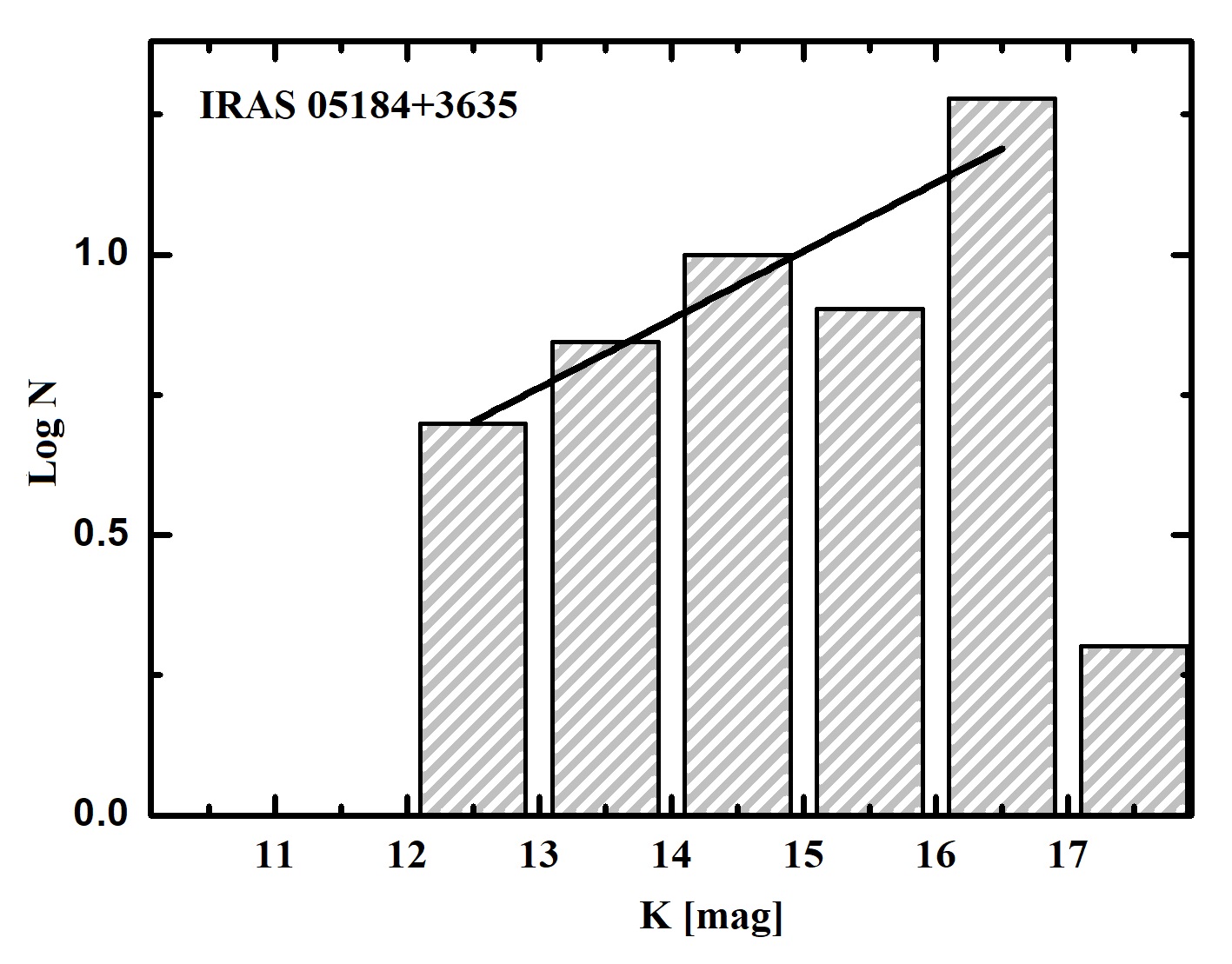}
\includegraphics[width=0.484\linewidth]{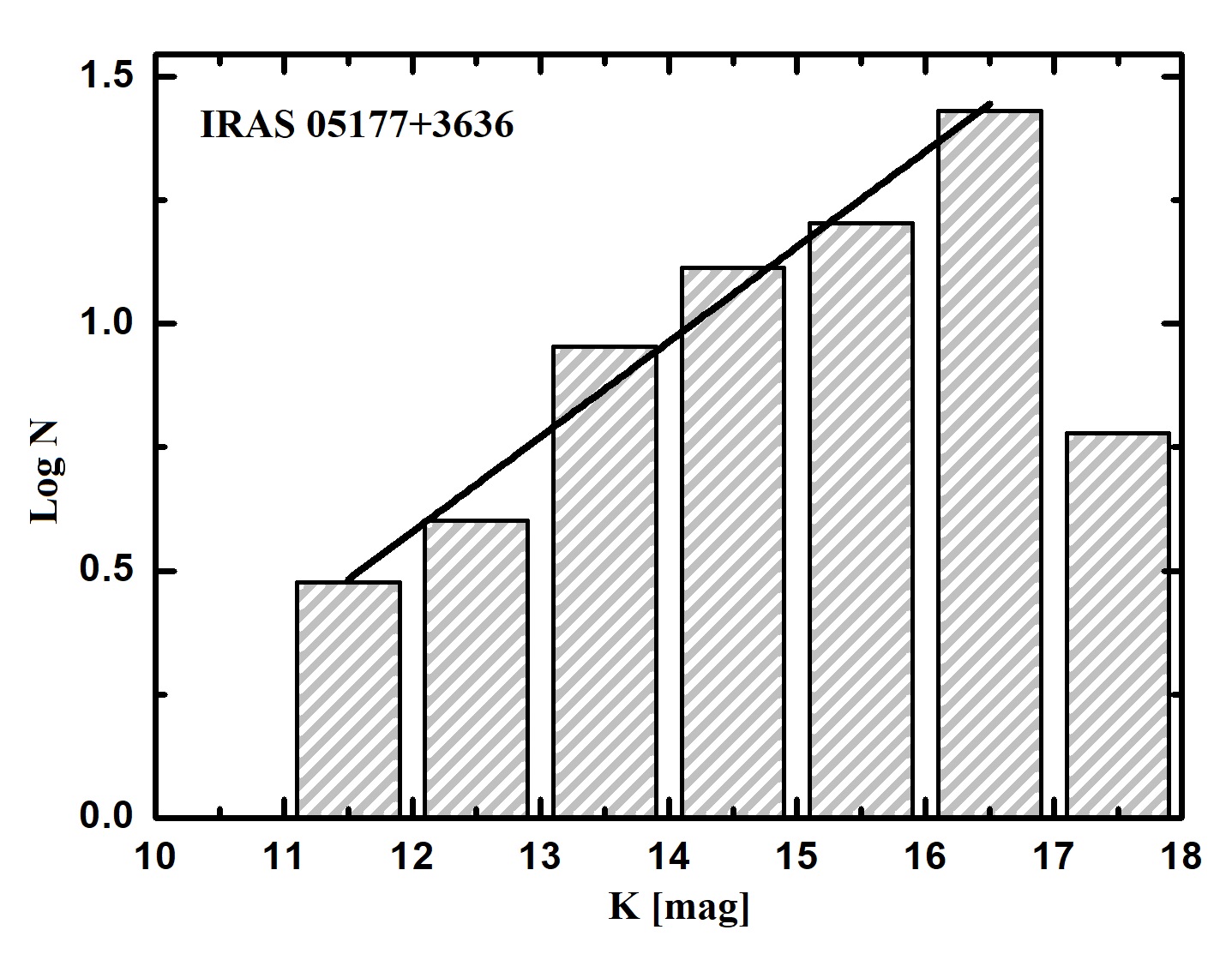}
\includegraphics[width=0.484\linewidth]{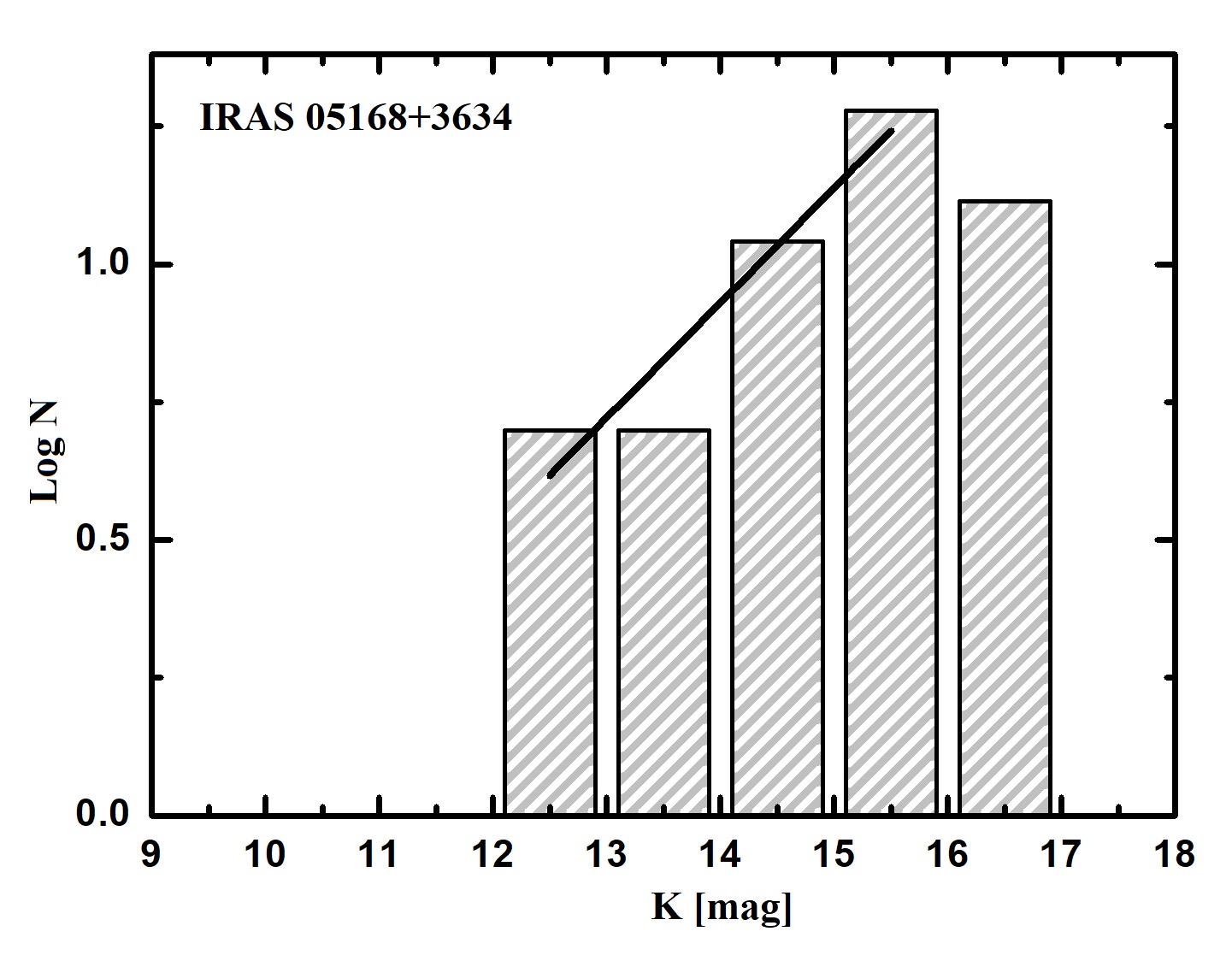}
\includegraphics[width=0.484\linewidth]{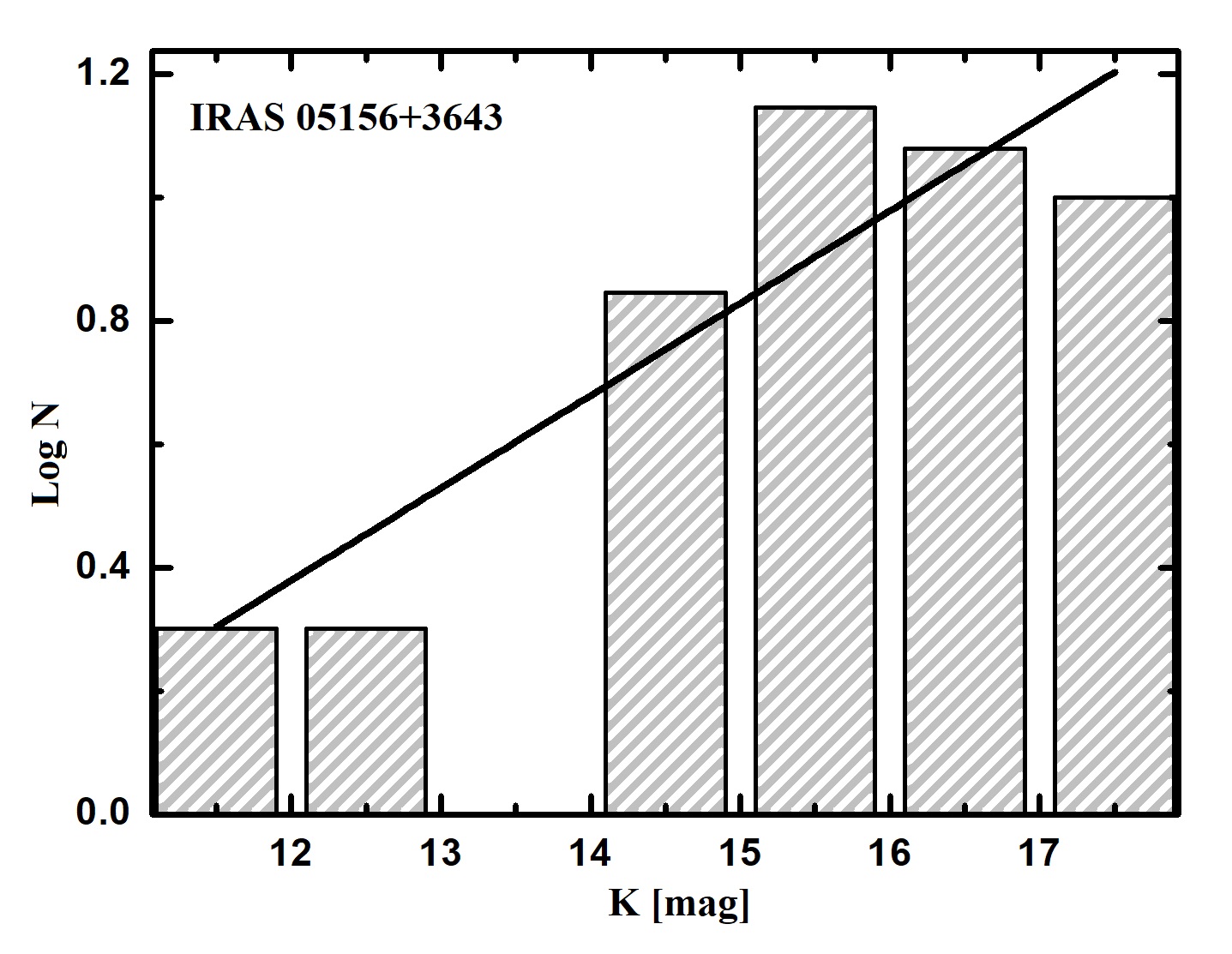}
\caption{ K luminosity functions derived for the subregions as a histogram of the number of stars in logarithm vs. apparent K magnitude. The bin size corresponds to 1\,mag. The linear fits are represented by the straight lines and the slopes obtained are given in each figure.}
\label{fig:6}
\end{figure*}

\subsection{K luminosity function}
\label{3.5}
The luminosity function in the K-band (KLF) is frequently used in studies of young clusters and star-forming regions as a diagnostic tool of the initial mass function (IMF) and the star formation history of their stellar populations \citep{zinnecker93,lada95,jose12}. Thus, the observed KLF is a result of the IMF, the PMS evolution, and the star-forming history. Due to the uncertainty in the selected YSO ages (see Figure \ref{fig:5}), it is impossible to derive an individual mass for each source and therefore the mass-luminosity relation cannot be constructed in order to get the IMF. Nevertheless, it is possible to construct the KLF to constrain the age of the embedded stellar population in each star-forming region independently. As pointed out by \citet{lada96}, the age of a subregion can be estimated by comparing its KLF to the observed KLFs of other young clusters, i.e., the KLF slope could be an age indicator of young clusters. The KLF can be defined as dN(K)/dK$\propto10^{\alpha K}$, where $\alpha$ is the slope of the power law and N(K) is the number of stars brighter than K\,mag. The KLF slope is estimated by fitting the number of YSOs in 1\,mag bins using a linear least-squares fitting routine. Since the slope is not affected by the intracluster extinction as long as the extinction is independent of the stellar mass \citep{megeath96}, we can use all the sources in the sample to define the slope without the complication of extinction correction. Therefore, Figure \ref{fig:6} shows the observed KLFs of YSOs detected in  four subregions separately. In the case of the IRAS\,05162+3639 subregion, there are not enough YSOs  to construct the KLF and so it is impossible to estimate the value of the $\alpha$ slope. The KLFs corresponding fitted slopes are provided in Table \ref{tab:4}, which are lower than the typical values reported for young embedded clusters \citep[$\alpha\,\sim$0.4; e.g.,][]{lada91,lada03,baug15} and within the errors, the KLFs for all four subregions seem to match each other. For clusters up to 10\,Myr old, the KLF slope gets steeper as the cluster gets older \citep{ali95,lada95}. There are many studies on the KLFs of young clusters. \citet{megeath96} found similar values of $\alpha$
slopes around the W3 IRS\,5 cluster ($\alpha$=0.24), and that it is consistent with an age of 0.3\,Myr with a Miller-Scalo IMF. \citet{ojha04} estimated the KLF slope of NGC\,7538 to be $\alpha$=0.28$\pm$0.02 and with an age of $\sim$1\,Myr.

In addition, according to the calculation of \citet{massi00}, $\alpha$ values between 0.2 and 0.28 are consistent with an age range of 0.1--3\,Myr. The $\alpha$ values of the observable subregions are close to this range of values; therefore, the age of all four subregions can be estimated between 0.1 and 3\,Myr, which also closely reflects the location of stellar objects relative to the isochrones. The KLF of the IRAS\,05156+3643 star-forming region differs from the other subregions, i.e., a sharp decline does not occur in the end, which means that in this case the KLF reflects the real picture of the subregion, and in the remaining cases the decline is abrupt, reflecting the completeness limit of the survey. The KLF of IRAS\,05156+3643 star-forming region complicates the resolution of the $\alpha$ slope value, but in any case it falls within the 0.15--0.28 range.

\subsection{SED analysis}
\label{3.6}
In order to learn about the evolutionary stages of the YSOs that have measured magnitudes in longer wavelengths, we constructed their spectral energy distributions (SEDs) and fitted them with the radiative transfer models of \citet{robitaille07}. The models assume an accretion scenario in the star formation process where a central star is surrounded by an accretion disk, an infalling flattened envelope, and the presence of bipolar cavities. We  used the command-line version of the SED fitting tool where a large number of precomputed models are available. The SEDs are constructed for 45 Class\,I and 75 Class\,II evolutionary stage YSOs. This procedure was done using wavelengths ranging from 1.1\,$\mu$m to 160\,$\mu$m, in particular J, H, and K (UKIDSS); 3.6 and 4.5\,$\mu$m ($Spitzer$ IRAC); 3.4, 4.6, 12, and 22\,$\mu$m (WISE); 8.28, 12.13, 14.65, and 21.3\,$\mu$m (MSX); 12 and 25\,$\mu$m (IRAS); and 70 and 160\,$\mu$m ($Herschel$ PACS). We took the apparent magnitudes of those surveys where the object was detected as a point source, i.e., we did not take data from the $Herschel$ PACS extended source list and SPIRE 250,350,500\,$\mu$m surveys (all YSOs in the molecular cloud were detected as extended in the $Herschel$ SPIRE survey). We also did not take the  IRAS 60 and 100\,$\mu$m data, but instead we took the $Herschel$ PACS Point Source Catalogs 70 and 160\,$\mu$m data, which provide better resolution than the IRAS 60 and 100\,$\mu$m. The SED fit was carried out using both distance estimations: 1.88 and 6.1\,kpc. We used the ranges of the interstellar extinction (A$_v$) and the distances of 1--40\,mag, and 5.5--6.5\,kpc and 1.6--2\,kpc, respectively.

\begin{figure*}
\centering
\includegraphics[width=0.32\linewidth]{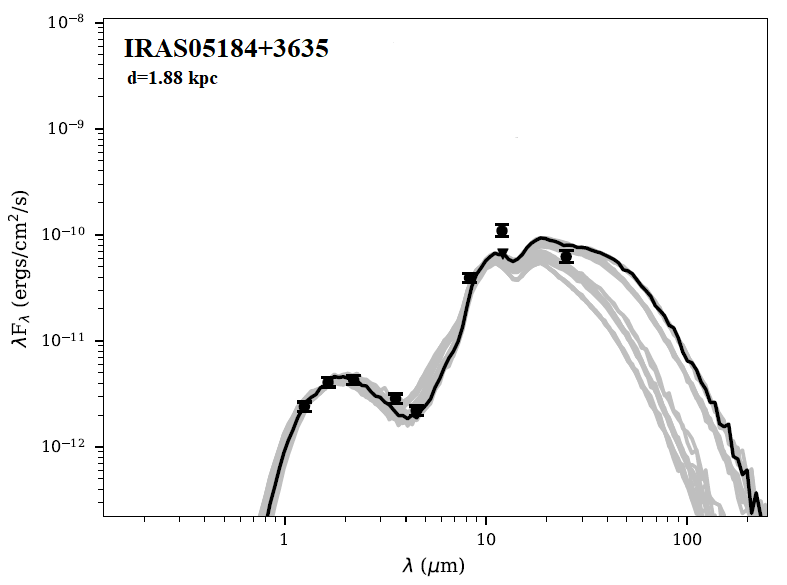}
\includegraphics[width=0.32\linewidth]{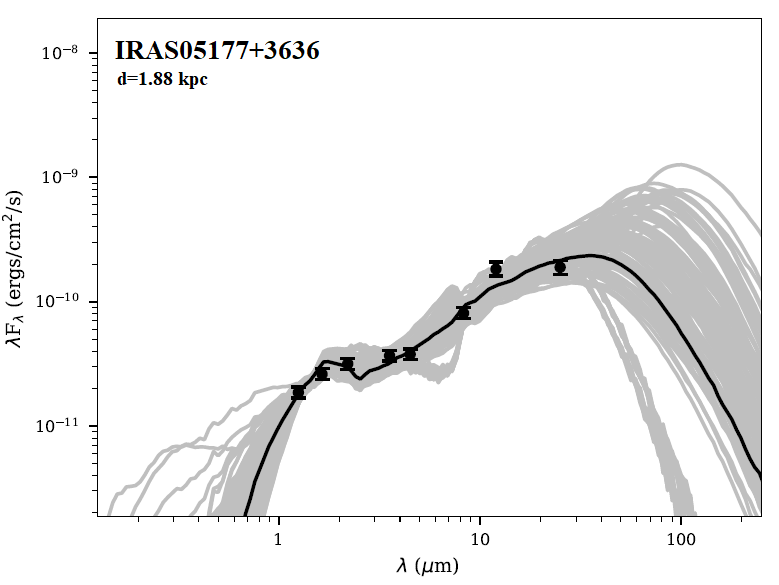}
\includegraphics[width=0.32\linewidth]{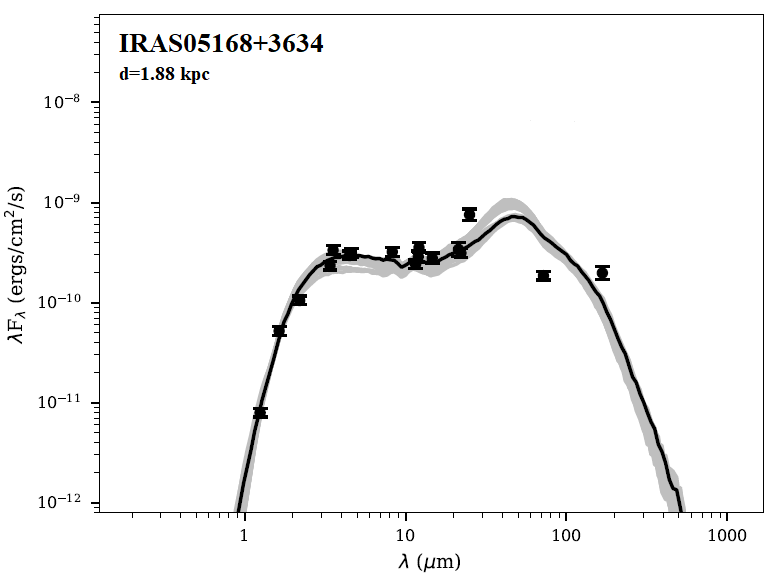}
\includegraphics[width=0.32\linewidth]{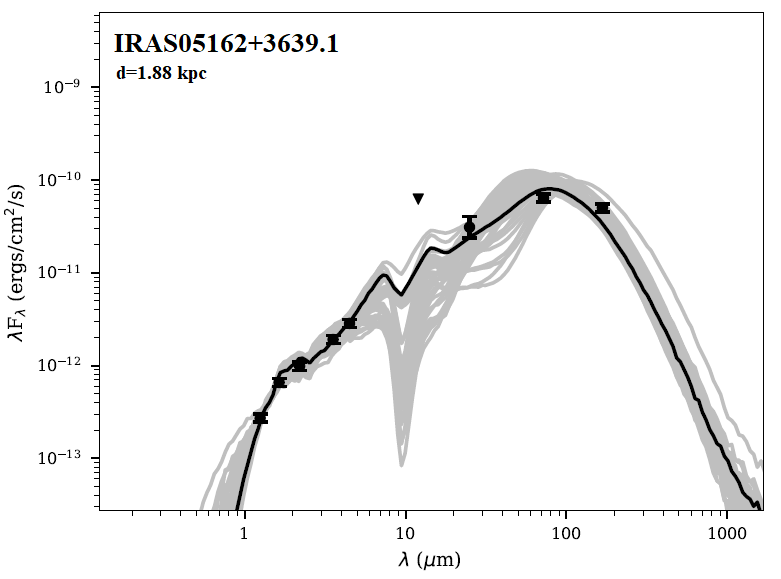}
\includegraphics[width=0.32\linewidth]{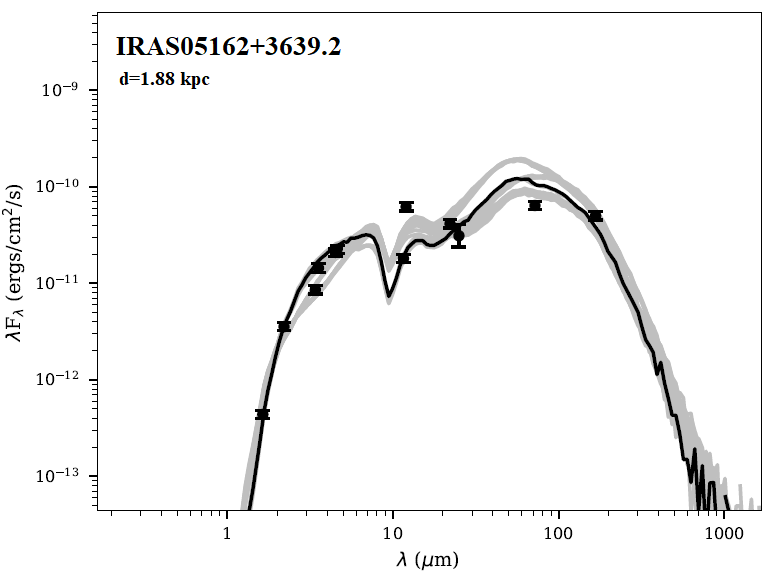}
\includegraphics[width=0.32\linewidth]{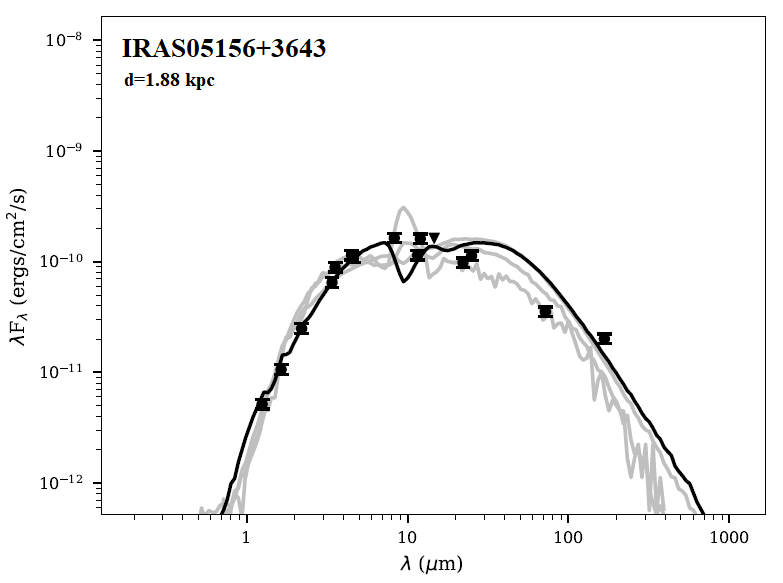}
\includegraphics[width=0.32\linewidth]{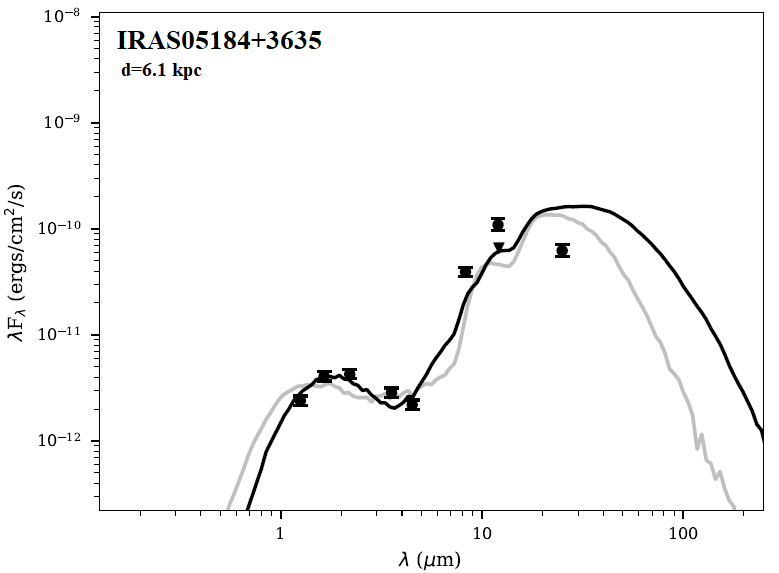}
\includegraphics[width=0.32\linewidth]{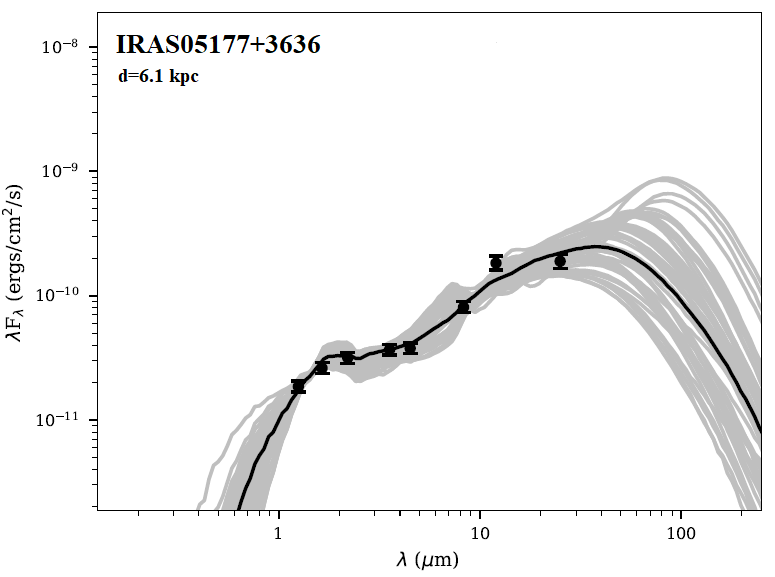}
\includegraphics[width=0.32\linewidth]{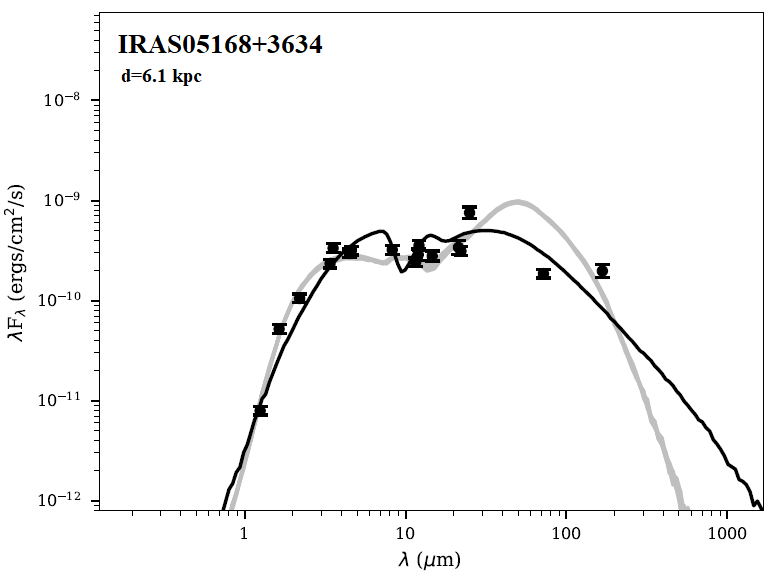}
\includegraphics[width=0.32\linewidth]{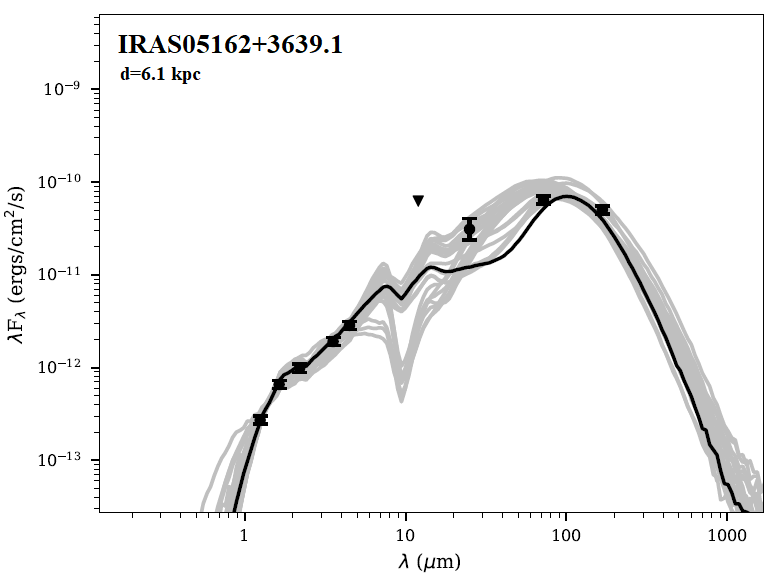}
\includegraphics[width=0.32\linewidth]{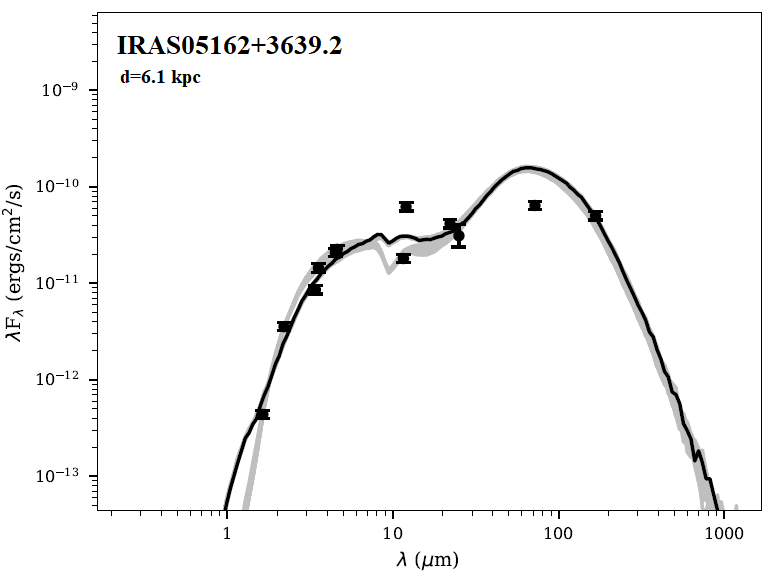}
\includegraphics[width=0.32\linewidth]{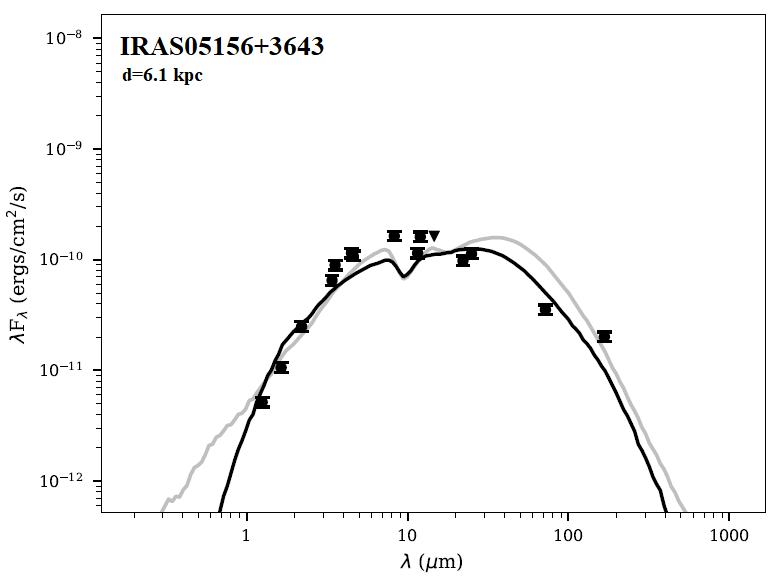}
\caption{Infrared SEDs of IRAS sources are fitted to the models of \citet{robitaille07} for distances of 1.88 and 6.1\,kpc. The filled circles represent the input fluxes, while triangles represent upper limits from the respective clumps. The black line shows the best fitting model and gray lines show the subsequent good fits.}
\label{fig:7}
\end{figure*}

Figure \ref{fig:7} shows the constructed SEDs for those objects  associated with IRAS sources in the near and far distances. To identify the representative values of different physical parameters, the tool retrieved the best fit model and all models for which the differences between their $\chi^2$ values and the best $\chi^2$ were smaller than 3N, where N is the number of the  data points used  \citep[as suggested in][]{robitaille07}. This approach was taken because the sampling of the model grid is too sparse to effectively determine the minima of the $\chi^2$ surface and consequently obtain the confidence intervals.
                 
Table \ref{tab:7} shows the weighted means and the standard deviations of parameters for all models with $\chi^2$$-$$\chi^{2}_{best}$ < 3N at near and far distances: 1.88 and 6.1\,kpc. In Table \ref{tab:7}, the members of subregions are separated from each other and the parameters of member of each subregion are listed in the following order: first, the received parameters of IRAS sources, then the parameters of MSX sources (if there is an identical object with any MSX source in that subregion). Next are the parameters of Class\,I objects, and finally those of  Class\,II. The numbering of the objects was done according to Table \ref{tab:5}.

According to the results of SED fitting tool at  distances of 1.88 and 6.1\,kpc, objects associated with IRAS and MSX sources can be classified as middle-mass YSOs (Table \ref{tab:7}), which confirms the results obtained in the color-color and color-magnitude diagrams. The results of the SED fitting tool in general confirm the age estimations obtained by the KLF slope for each subregion. The ratio between disk accretion and envelope infall rates of our objects  correspond closely to those parameters given for Class\,I objects in \citet{grave09}.

\subsection{Substructures of molecular cloud}
\label{3.7}
\begin{figure*}
\centering
\includegraphics[width=0.46\linewidth]{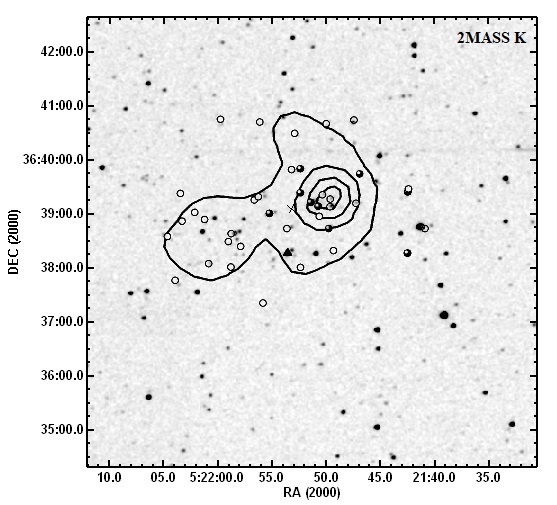}
\includegraphics[width=0.46\linewidth]{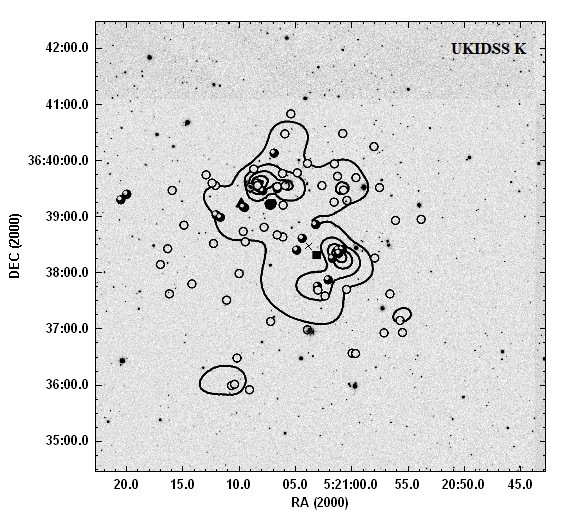}
\includegraphics[width=0.46\linewidth]{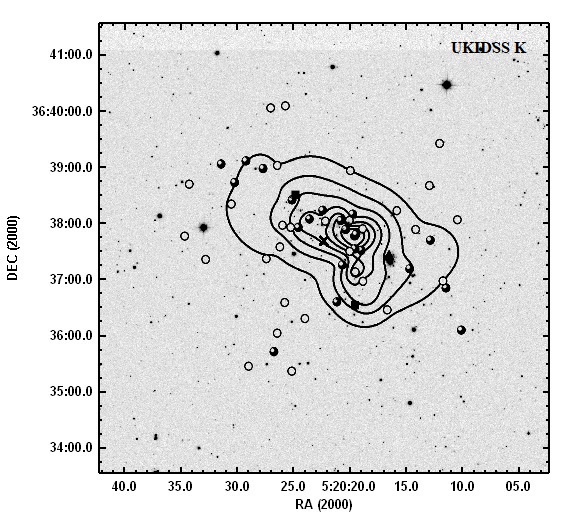}
\includegraphics[width=0.46\linewidth]{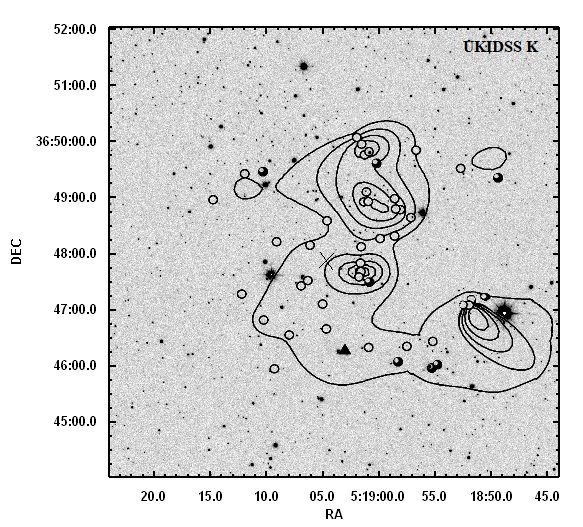}
\includegraphics[width=0.46\linewidth]{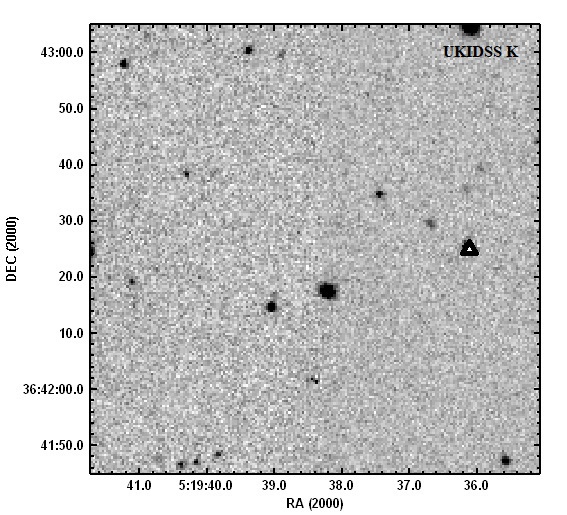}
\caption{LD distribution of the YSOs and spatial distribution of the Class\,I and Class\,II objects superposed on the K-band images. Filled and open circles  represent the Class\,I and Class\,II objects, respectively. The triangles and squares show the location of IRAS and MSX sources, respectively, which are also classified as Class\,I objects. The geometric center of each subregion is indicated by a cross.}
\label{fig:8}
\end{figure*}

In this section, we give an overview of the substructures of four clearly visible subregions found in the molecular cloud. Figure \ref{fig:8} shows the local density (LD) distribution of selected YSOs (N = 240) within the obtained radii (see Section \ref{3.3}) on the K-band images. A LD has been decided for each object on the surface with a radius equal to the distance to the closest n-th star. The value of LD for the first isodense exceeds the density of a field with a value of 3$\sigma$. We investigate separately the substructures of subregions, and the distributions of Class\,I and Class\,II evolutionary stage objects in the subregions. Each of the subregions is an embedded cluster.

The LD distribution of all the YSOs in the subregions shows elongated structures. The elongated structure was also evident in some of the earlier studies of clusters \citep{carpenter97,hillenbrand98}. The elongation appears to be a result of the primordial structure in the cloud, and the elongation is aligned with the filamentary structure seen in the parent molecular cloud \citep{allen07} and they are often composed of subregions \citep{lada96,chen97,megeath97,allen02}. All subregions have an expressed elongation and each consists of subgroups.

Figure \ref{fig:8} (top left panel) shows the LD distribution of YSOs in the IRAS\,05184+3635 star-forming region. The IRAS\,05184+3635 star-forming region has a bimodal structure and the main concentration of objects is located $\backsim$40\,arcsec northwest (NW) of IRAS\,05184+3635, which is associated with a MSX object (G170.8276+00.0098).

The LD distribution of YSOs in the IRAS\,05177+3636 star-forming region is shown in Figure \ref{fig:8} (top right panel). The IRAS\,05177+3636 star-forming region also has a bimodal structure and consists of two main subgroups, each  with an elongation. There are three MSX sources in the IRAS\,05177+3636 star-forming region, one of them (G170.7268-00.1012) is associated with IRAS\,05177+3636. The second MSX source (G170.7196-00.1118) is located in the northeast (NE) subgroup with IRAS\,05177+3636, and the third (G170.7247-00.1388) is in the second subgroup $\backsim$2.2\,arcmin to the southwest (SW) of IRAS\,05177+3636. However, it should be noted that  IRAS\,05177+3636 and all the MSX sources are on the edges of two subgroups.

The LD distribution in IRAS\,05168+3634 star-forming region demonstrates unique elongation (Figure \ref{fig:8}, bottom left panel). There are no clearly visible subgroups, but three poles associated with MSX sources can be seen. One of the subgroups is located around  IRAS\,05168+3634, which is the dominant object in the given subgroup. IRAS\,05168+3634 is identified with MSX\,G170.6575-00.2685. Another subgroup is found to the NE  of  IRAS\,05168+3634. \citet{molinari98} detected 6 cm radio emission 102\,arcsec away from  IRAS\,05168+3634 whose coordinates are $\alpha_{2000}$=05:20:23.53, $\delta_{2000}$=+36:38:17.58. 

The location of the second MSX\,G170.6589-00.2334 object is $\sim$2.118\,arcmin NE of the position of  IRAS\,05168+3634. The third subgroup in the IRAS\,05168+3634 star-forming region is located to the southeast (SE)  of the IRAS\,05168+3634. 

The MSX\,G170.6758-00.2691 object in the IRAS\,05168+3634 star-forming region is $\sim$1.135\,arcmin SE of the position of  IRAS\,05168+3634  and is located in the subgroup. In the case of the IRAS\,05168+3634 subregion the IRAS and MSX sources are also located in the edges of subgroups.

The IRAS\,05156+3643 star-forming region has the most complex structure (Figure \ref{fig:8}, bottom right panel). It consists of three subgroups that show elongated structures, and  one of them is even composed of two smaller subgroups. IRAS\,05156+3643 is out of these three subgroups. This IRAS object is associated with MSX\,G170.3964-00.3827. 

The LD distribution on the IRAS\,05162+3639 star-forming region cannot provide any information because there are only five objects. It should be noted that all of these five objects were classified as Class I evolutionary stage objects (see text above), and that two of them were identified with IRAS\,05162+3639. This binary needs a separate study and it should be noted that the results obtained from the SED are likely to be incorrect.

If the observed morphologies of embedded clusters result from the filamentary and clumpy nature of the parental molecular clouds, then the younger Class\,0/I objects, which have the least time to move away from their star formation sites, should show more pronounced structures than the older, PMS Class\,II and Class\,III stellar objects. However, there is no  difference between the distributions of Class\,I and II objects in the field of the subregions, but there is a big difference between the distributions of Class\,I and II objects in the molecular cloud as a whole (Figure \ref{fig:3}).

\section{Conclusion}
\label{4}

Our investigations show that the IRAS\,05168+3634 star-forming region, considered in previous studies, has a more complicated structure in the FIR wavelengths. Studies in FIR wavelengths suggest that the IRAS\,05168+3634 star-forming region is located in a quite large molecular cloud within a region of approximately 24\,arcmin radius (the center of the molecular cloud is conditionally selected as IRAS\,05168+3634), which in turn consists of four additional star-forming regions. From the statistical analysis it follows that all IRAS star-forming regions are at the same distance: 6.1\,kpc \citep{molinari96} or 1.88\,kpc \citep{sakai12}, i.e., there  is a small probability that these star-forming regions may be the result of superposition. In addition, these regions repeat the shape of the molecular cloud. Also the distances of IRAS\,05184+3635 and IRAS\,05177+3636 assessed based on the $^{13}$CO velocities were evaluated at the same 1.4\,kpc value, which  coincides with the distance of IRAS\,05168+3634 based on trigonometric parallax.

In this paper, we also analyzed the stellar content of the molecular cloud and separately for each star-forming region associated with IRAS sources using the J, H, K UKIDSS-DR6; [3.6], [4.5]\,$\mu$m $Spitzer$ GLIMPSE\,360; and [3.4], [4.6], [12], [22]\,$\mu$m ALLWISE survey images and photometric data to construct color-color and color-magnitude diagrams, which are useful tools for identifying YSO candidates. Using NIR and MIR photometric data, we obtain the census of the young stellar populations and their characteristics within a 24\,arcmin radius region surrounding the molecular cloud, which includes 1224 candidate YSOs, 240 of which are concentrated around five IRAS sources. We classified 71 YSOs as evolutionary stage Class\,I objects and 132--169 YSOs (depending on the distance) as evolutionary stage Class\,II objects within the radii of subregions. One can see, depending on the color-color and color-magnitude diagrams, the number of Class\,II evolutionary stage objects is changing, but the number of Class\,I evolutionary stage objects is not changing, and the conclusion is that the selection of Class\,I objects is highly accurate. It should be noted that, unlike the Class\,II objects, the Class\,I objects are located mainly in the filaments of the molecular cloud, i.e., the distribution of Class\,I objects reflects the initial state of the parent molecular cloud. The estimated distances and the interstellar extinctions of each subregion were taken into account in the corrections of J and K magnitudes for the color-magnitude diagrams, which generates a great difference in the number of Class\,II objects. From the color-magnitude diagram analysis using evolutionary models of various ages by \citet{siess00}, there is a large distribution of ages, but the age of most of them is on the order of 10$^5$  years. A mass completeness limit has been quoted for two distances. Selected YSOs located in the IRAS\,05184+3635, IRAS\,05168+3634, and IRAS\,05156+3643 star-forming regions are explicitly targeted at a specific mass range for two distances;  for objects in the IRAS\,05177+3636 star-forming region there is no similar dependency. The members of the IRAS\,05184+3635 and IRAS\,05156+3643 subregions are targeted to a lower mass range, i.e., the members of IRAS\,05184+3635 are distributed in the ranges of 0.3--1.5 and 0.8--5 M$_{sun}$  and the members of IRAS\,05156+3643 in the ranges 0.2--1.6 and 0.6-3 M$_{sun}$  at distances of  1.88 and 6.1\,kpc, respectively. Conversely, the members of the IRAS\,05168+3634 subregion are distributed in a higher mass range, i.e., they are at 0.5--2.5 and 0.9--$>$7 M$_{sun}$  at distances of  1.88 and 6.1\,kpc, respectively.

We have also calculated the slope of the KLFs for four subregions, namely IRAS\,05184+3635, IRAS\,05177+3636, IRAS\,05168+3634, and IRAS\,05156+3643. The KLF of these subregions shows  unusually low values for the $\alpha$ slope: 0.12--0.21. According to the values of the slopes of the KLFs, the age of all four subregions can be estimated between 0.1 and 3\,Myr. There are not enough YSOs within the radius of the IRAS\,05162+3639 subregion, so the KLFs for this subregion were not constructed.

The SEDs were constructed for 45 and 75 YSOs with evolutionary stages  Class\,I and Class\,II, respectively. This procedure was done using wavelengths ranging from 1.1\,$\mu$m to 160\,$\mu$m. According to the results of the SED fitting tool,  IRAS\,05184+3635, IRAS\,05177+3636, and IRAS\,05162+3639  can be classified as Class\,I for both  distances. The sources IRAS\,05168+3634 and IRAS\,05156+3643 can be classified as flat-spectrum objects for both distances. At the distance of 6.1\,kpc, all IRAS sources have 6--7 M$_{sun}$, except for IRAS\,05156+3643, and one of the two objects associated with the IRAS\,05162+3639 binary object (both have 3 M$_{sun}$ mass). At the distance of 1.88\,kpc, estimated masses vary considerably only in the case of IRAS\,05184+3635 and IRAS\,05177+3636, and the mass estimation is the highest for IRAS\,05168+3634: 5 M$_{sun}$. The results of the SED fitting tool, in general, are well correlated with the age estimations obtained by the KLF slope for each subregion. According to the results of the SED fitting tool, at  distances of 1.88 and 6.1\,kpc, IRAS and MSX sources can be classified as middle-mass YSOs, which confirms the results obtained in the color-color and color-magnitude diagrams. 

All star-forming regions have complicated structures. They contained the subgroups, which are mainly associated with IRAS and MSX sources. The observed young subregions and parental molecular cloud morphologies are similar and elongated. This is particularly well expressed when only the youngest Class\,I/0 sources are considered. Perhaps the similarities in  morphologies  result from the distribution of fragmentation sites in the parent cloud. The Class\,I sources are often distributed along filamentary structures, while the Class\,II sources are more widely distributed. A similar distribution of Class\,II evolutionary stage objects can be explained by the fact that these objects have had enough time to leave their own birthplaces. In the case of Class\,I evolutionary stage objects, that time is small, and this is why they are located on the filaments of parent molecular cloud, thus they clearly  reflect the initial state of the parent cloud.

\begin{acknowledgements}
\small
I am very grateful to the anonymous referee for the helpful comments and suggestions. I thank my supervisor Dr. Elena Nikoghosyan for her support. This research has made use the data obtained at UKIRT, which is supported by NASA and operated under an agreement among the University of Hawaii, the University of Arizona, and Lockheed Martin Advanced Technology Center; operations are enabled through the cooperation of the East Asian Observatory. We gratefully acknowledge the use of data from the NASA/IPAC Infrared Science Archive, which is operated by the Jet Propulsion Laboratory, California Institute of Technology, under contract with the National Aeronautics and Space Administration. I thank my colleagues in the GLIMPSE\,360 $Spitzer$ Legacy Surveys. This work was made possible by a research grant from the Armenian National Science and Education Fund (ANSEF) based in New York, USA.
\end{acknowledgements}

\bibliographystyle{aa}
\bibliography{iras05}

\newgeometry{margin=1.2cm} 
\begin{landscape}
{\fontsize{8}{7}\selectfont 
\begin{center}
\begin{longtable}{l l l *{9}{c} l}
\caption[]{Members of subregions with NIR and MIR photometry} \label{tab:5} \\
\hline\hline\noalign{\smallskip}
ID & $\alpha$(2000) & $\delta$(2000)  & J   &  H &  K   & [3.6]\,$\mu$m    & [4.5]\,$\mu$m  & W1  &  W2   &  W3 &     W4  &       Final/class     \\
        &       (hh mm ss)                      &       (dd mm ss)                      &               (mag)           &               (mag)           &               (mag)           &               (mag)           &               (mag)           &               (mag)           &               (mag)           &               (mag)           &               (mag)           &               \\
\hline\noalign{\smallskip}
(1)     &               (2)             &               (3)                 &            (4)             &               (5)             &               (6)             &               (7)             &               (8)             &               (9)             &               (10)            &               (11)            &               (12)            &       (13)    \\
\hline \noalign{\smallskip}
\multicolumn{13}{c}{IRAS\,05184+3635}\\
\hline \noalign{\smallskip}
1       &       5       21      41.35   &       36      38      46.50   &       10.333  $\pm$   0.000   &       10.642  $\pm$   0.000   &       10.076  $\pm$   0.000   &       9.595   $\pm$   0.043   &       9.433   $\pm$   0.040   &       10.109  $\pm$   0.021   &       9.957   $\pm$   0.018   &       9.098   $\pm$   0.032   &       8.375   $\pm$   0.294   &       Class\,II \\
2\tablefootmark{*}      &       5       21      42.79   &       36      39      29.08   &       17.751  $\pm$   0.036   &       17.182  $\pm$   0.039   &       16.825  $\pm$   0.056   &       16.146  $\pm$   0.101   &       15.999  $\pm$   0.109   &               $-$             &               $-$             &               $-$             &               $-$             &       Class\,II \\
\hline
\end{longtable}
\end{center}
}
\tablefoot{
(1)-ID number in final list of sub-region members, (2),(3)-The position is taken from the UKIDSS survey, (4)-(12)-apparent magnitudes with errors, (13)-Classification of YSOs according to color-color diagrams W1-W4 are four WISE survey bands, \tablefoottext{*}{- objects that are located to the left of ZAMS in one color - magnitude diagram,} \tablefoottext{**}{- objects that are located to the left of ZAMS in both color - magnitude diagrams}
}
\end{landscape}
\restoregeometry

\newgeometry{margin=1.2cm} 
\begin{landscape}
{\fontsize{7}{7}\selectfont 
\begin{center}
\begin{longtable}{l l *{8}{c} l l *{3}{c}}
\caption[]{Members of subregions with FIR photometry} \label{tab:6} \\
\hline \hline
\noalign{\smallskip}
ID      &       Name    &       [8.28]\,$\mu$m                       &       [12.13]\,$\mu$m                      &       [14.65]\,$\mu$m                     &        [21.3]\,$\mu$m                       &        [12]\,$\mu$m                       &        [25]\,$\mu$m                         &        [60]\,$\mu$m                        &        [100]\,$\mu$m                        &        [70]\,$\mu$m                        &        [160]\,$\mu$m                        &        [250]\,$\mu$m                       &        [350]\,$\mu$m                        &        [500]\,$\mu$m                       \\
\noalign{\smallskip}
        &               &               (mJy)           &               (mJy)           &               (mJy)           &               (mJy)           &               (mJy)           &               (mJy)           &               (mJy)           &               (mJy)           &               (mJy)           &               (mJy)           &               (mJy)           &               (mJy)           &               (mJy)           \\
\hline
\noalign{\smallskip}
(1)     &       (2)     &               (3)             &               (4)             &               (5)             &               (6)             &               (7)             &               (8)             &               (9)             &               (10)            &               (11)            &               (12)            &               (13)            &               (14)            &               (15)            \\
\noalign{\smallskip}
\hline \noalign{\smallskip}
15      &               &                               &                               &                               &                               &                               &                               &                               &                               &       391     $\pm$   64      &                               &                               &                               &                               \\
31      &       IRAS\,05184+3635  &       109     $\pm$   11      &       266     $\pm$   53      &                               &                               &       440     $\pm$   57      &       523     $\pm$   68      &       5800    $\pm$   638     &       19500   $\pm$   5070    &       1415    $\pm$   10.552\tablefootmark{*}      &                               &                               &                               &                               \\
51      &               &                               &                               &                               &                               &                               &                               &                               &                               &       928     $\pm$   152     &       6942    $\pm$   5.215\tablefootmark{**}        &                               &                               &                               \\
60      &               &                               &                               &                               &                               &                               &                               &                               &                               &                               &       1059    $\pm$   94      &                               &                               &                               \\
85      &               &                               &                               &                               &                               &                               &                               &                               &                               &       1006    $\pm$   103     &       4069    $\pm$   671     &                               &                               &                               \\
90      &               &                               &                               &                               &                               &                               &                               &                               &                               &                               &       130     $\pm$   41      &                               &                               &                               \\
92      &               &                               &                               &                               &                               &                               &                               &                               &                               &       2322    $\pm$   20      &       3421    $\pm$   110     &                               &                               &                               \\
95      &               &                               &                               &                               &                               &                               &                               &                               &                               &                               &       130     $\pm$   41      &                               &                               &                               \\
96      &               &                               &                               &                               &                               &                               &                               &                               &                               &       1006    $\pm$   103     &       4069    $\pm$   671     &                               &                               &                               \\
97      &       IRAS\,05177+3636  &       224     $\pm$   22      &                               &                               &                               &       738     $\pm$   52      &       1590    $\pm$   127     &       4290    $\pm$   987     &       15900   $\pm$   2703    &                               &                               &                               &                               &                               \\
101     &       G170.7247-00.1388       &       171  $\pm$  20      &                               &                               &                               &                               &                               &                               &                               &       814     $\pm$   125     &       1444    $\pm$   979     &                               &                               &                               \\
112     &               &                               &                               &                               &                               &                               &                               &                               &                               &       2951    $\pm$   34      &       7689    $\pm$   279     &                               &                               &                               \\
126     &               &                               &                               &                               &                               &                               &                               &                               &                               &       2951    $\pm$   34      &       7689    $\pm$   279     &                               &                               &                               \\
128     &       G170.7196-00.1118       &       181     $\pm$   21      &                               &                               &                               &                               &                               &                               &                               &       1261    $\pm$   58      &       3683    $\pm$   349     &                               &                               &                               \\
130     &               &                               &                               &                               &                               &                               &                               &                               &                               &       2693    $\pm$   70      &       5185    $\pm$   594     &                               &                               &                               \\
177     &       G170.6589-00.2334       &       159     $\pm$   19      &                               &       859     $\pm$   143     &                               &                               &                               &                               &                               &       5163    $\pm$   813     &                               &                               &                               &                               \\
178     &               &                               &                               &                               &                               &                               &                               &                               &                               &                               &       291     $\pm$   47      &                               &                               &                               \\
182     &       G170.6758-00.2691       &       185     $\pm$   22      &       662     $\pm$   132     &       525     $\pm$   105     &                               &                               &                               &                               &                               &       1735    $\pm$   75      &       5052    $\pm$   7.704\tablefootmark{**}        &                               &                               &                               \\
183     &       IRAS\,05168+3634  &       891     $\pm$   89      &       1450    $\pm$   171     &       1370    $\pm$   161     &       2450    $\pm$   408     &       1160    $\pm$   81      &       6340    $\pm$   444     &       167000  $\pm$   23380   &       379000  $\pm$   45480   &       4472    $\pm$   103     &       11125   $\pm$   1724    &                               &                               &                               \\
189     &               &                               &                               &                               &                               &                               &                               &                               &                               &       109     $\pm$   32      &                               &                               &                               &                               \\
190     &       IRAS\,05162+3639  &                               &                               &                               &                               &       250     $\pm$   125     &       269     $\pm$   73      &       1280    $\pm$   179     &       4590    $\pm$   872     &       1545    $\pm$   12      &       2814    $\pm$   74      &                               &                               &                               \\
192     &       IRAS\,05162+3639  &                               &                               &                               &                               &       250     $\pm$   125     &       269     $\pm$   73      &       1280    $\pm$   179     &       4590    $\pm$   872     &       1545    $\pm$   12      &       2814    $\pm$   74      &                               &                               &                               \\
216     &               &                               &                               &                               &                               &                               &                               &                               &                               &                               &                               &       7684    $\pm$   212     &       4171.5  $\pm$   111     &       2295    $\pm$   72      \\
219     &               &                               &                               &                               &                               &                               &                               &                               &                               &       270     $\pm$   22      &       574     $\pm$   60      &       7684    $\pm$   212     &       4171.5  $\pm$   111     &       2295    $\pm$   72      \\
229     &               &                               &                               &                               &                               &                               &                               &                               &                               &                               &                               &       7684    $\pm$   212     &       4171.5  $\pm$   111     &       2295    $\pm$   72      \\
232     &               &                               &                               &                               &                               &                               &                               &                               &                               &       185     $\pm$   12      &                               &                               &                               &                               \\
234     &       IRAS\,05156+3643  &       455     $\pm$   46      &                               &       797     $\pm$   159     &                               &       649     $\pm$   58      &       947     $\pm$   85      &       1810    $\pm$   181     &       3020    $\pm$   332     &       853     $\pm$   18      &       1128    $\pm$   66      &       2031    $\pm$   119     &                               &                               \\
238     &               &                               &                               &                               &                               &                               &                               &                               &                               &                               &                               &                               &       7098.4  $\pm$   126     &       5575    $\pm$   105     \\
239     &               &                               &                               &                               &                               &                               &                               &                               &                               &       892     $\pm$   29      &       2668    $\pm$   136     &       6201    $\pm$   102     &       7098.4  $\pm$   126     &       5575    $\pm$   105     \\
240     &               &                               &                               &                               &                               &                               &                               &                               &                               &       144     $\pm$   43      &       932     $\pm$   149     &                               &                               &                               \\

\hline
\end{longtable}
\end{center}
}
\tablefoot{
(1) - ID number of sub-regions members taken from Table \ref{tab:5}, (2)- Names of IRAS and MSX sources, (3)-(6)- measured fluxes with errors in MSX catalog, (7)-10)-measured fluxes with errors in IRAS catalog, (11)-(12)- measured fluxes with errors in $Herschel$ PACS 70 and 160\,$\mu$m catalogs, (13)-(15)-measured fluxes with errors in $Herschel$ SPIRE 250, 350 and 500\,$\mu$m catalogs, \tablefoottext{*}{-fluxes taken from $Herschel$ PACS: Extended source list catalog in blue band (70\,$\mu$m),}
\tablefoottext{**}{- fluxes taken from $Herschel$ PACS: Extended source list catalog in red band (160\,$\mu$m)}
}
\end{landscape}
\restoregeometry

\newgeometry{margin=1.2cm} 
\begin{landscape}
{\fontsize{8}{6.5}\selectfont 
\begin{center}
\begin{longtable}{l *{17}{c}}
\caption[]{Parameters derived from \citet{robitaille07} models SED fitting for two distances.} \label{tab:7} \\
\hline \hline
\noalign{\smallskip}
ID   &   \multicolumn{2}{c}{A$_v$}       &    \multicolumn{2}{c}{Stellar age}     &   \multicolumn{2}{c}{Stellar mass}   &    \multicolumn{2}{c}{Temperature} &    \multicolumn{2}{c}{MdotE}  &  \multicolumn{2}{c}{MdotD}     & \multicolumn{2}{c}{L$_{Total}$} & \multicolumn{2}{c}{N$_{fit}$}   &        N$_{data}$  \\ 
&  \multicolumn{2}{c}{(mag)} & \multicolumn{2}{c}{(Log)(yr)} & \multicolumn{2}{c}{(M$_{\odot}$)} & \multicolumn{2}{c}{(Log)(K)} & \multicolumn{2}{c}{(Log)(M$_{\odot} yr^{-1}$)} & \multicolumn{2}{c}{(Log)(M$_{\odot} yr^{-1}$)} & \multicolumn{2}{c}{(Log)(L$_{\odot}$)} & & &\\
\noalign{\smallskip}
& 1.88   & 6.1    &    1.88         &       6.1  &                           1.88           &                 6.1        &                     1.88               &             6.1         &                    1.88   &                         6.1    &                         1.88       &                     6.1   &                          1.88         &                   6.1  &                           1.88    &        6.1     &\\      
\hline\noalign{\smallskip}
(1)   &                  (2)      &                       (3) &                            (4)    &                         (5)  &                           (6)  &                           (7)        &                     (8) &                            (9)      &                       (10)                     &       (11)  &                          (12)   &                         (13)&                            (14)  &                          (15)    &                (16)     &       (17)    &        (18)    \\
\hline\noalign{\smallskip}
\multicolumn{18}{c}{IRAS\,05184+3635}\\
\hline
\end{longtable}

\end{center}
}
\tablefoot{
(1)-ID number of sub-regions members taken from Table \ref{tab:5}, (2)-(15)-are the weighted means and the standard deviations of parameters obtained by fitting tool for all models with best $\chi^{2}-\chi^{2}_{best}$ < 3N at two distances: 1.88 and 6.1\,kpc, (16),(17)-are number of fitting models at two distances: 1.88 and 6.1\,kpc, (18)-is the number of input fluxes.
}

\end{landscape}
\restoregeometry

\end{document}